\documentclass[useAMS,usenatbib,usegraphicx]{mn2e}

\title{Pressure from dark matter annihilation and the rotation curve of spiral
galaxies}

\author[M.~Wechakama and Y.~Ascasibar]
{
M.~Wechakama$^{1\star}$ and Y.~Ascasibar$^{2\dagger}$\\
$^1$Astrophysikalisches Institut Potsdam, An der Sternwarte 16, Potsdam 14482
(Germany)\\
$^2$Departamento de F\'\i sica Te\'{o}rica, Universidad Aut\'onoma de Madrid,
Madrid 28049 (Spain)
}

\date{MNRAS, accepted (31 December 2010)}

\newcommand{\be}{\begin{equation}}
\newcommand{\ee}{\end{equation}}
\newcommand{\bea}{\begin{eqnarray}}
\newcommand{\eea}{\end{eqnarray}}

\newcommand{\dd}{{\rm d}}
\newcommand{\deriv}[2]{\frac{\dd#1}{\dd#2}}

\usepackage{color}
\newcommand{\Referee}[1]{{\color{black}#1}}

\begin{document}

\maketitle

\begin{abstract}
\Referee
{
The rotation curves of spiral galaxies are one of the basic predictions of the
cold dark matter paradigm, and their shape in the innermost regions has been
hotly debated over the last decades.
The present work shows that dark matter annihilation into electron-positron
pairs may affect the observed rotation curve by a significant amount.
}
We adopt a
model-independent approach, where all the electrons and positrons are injected
with the same initial energy $E_0 \sim m_{\rm dm} c^2$ in the
range from 1~MeV to 1~TeV and the injection rate is constrained by INTEGRAL,
Fermi, and HESS data. The pressure of the relativistic
electron-positron gas is determined by solving the diffusion-loss equation,
considering inverse Compton scattering, synchrotron radiation,
Coulomb collisions, bremsstrahlung, and ionization. For values of the gas
density and magnetic field that are representative of the Milky Way,
it is estimated that pressure gradients are strong enough to balance gravity in
the central parts if $E_0<1$~GeV. The exact value depends
somewhat on the astrophysical parameters, and it changes dramatically with the
slope of the dark matter density profile. For very steep slopes,
as those expected from adiabatic contraction, the rotation curves of spiral
galaxies would be affected on $\sim$~kpc scales for most values of
$E_0$.
\Referee
{
By comparing the predicted rotation curves with observations of dwarf and low
surface brightness galaxies, we show that the pressure from dark matter
annihilation may improve the agreement between theory and observations in some
cases, but it also imposes severe constraints on the model parameters (most
notably, the inner slope of halo density profile, as well as the mass and the
annihilation cross-section of dark matter particles into electron-positron
pairs).
}
\end{abstract}


\footnotetext[1]{E-mail: maneenate@aip.de}
\footnotetext[2]{E-mail: yago.ascasibar@uam.es}

\pagestyle{myheadings}

\markboth{M.~Wechakama and Y.~Ascasibar}{Pressure from DM annihilation and
rotation curves }

%--------------------------------------------------------------------------
 \section{Introduction}
%--------------------------------------------------------------------------

Although cosmological observations are providing increasingly
convincing evidence that non-baryonic cold dark matter (CDM) is
the building block of structures in the Universe, the exact nature
of dark matter remains a mystery. A promising approach to the
problem relies on the possibility that dark matter particles
annihilate into observable products such as photons, neutrinos,
protons, anti-protons, electrons, and positrons. Thus one can aim
for indirect dark matter detection by looking for signatures of
the annihilation products \citep[see e.g.][]{Bertone+05}.

The most commonly studied signature is the emission of gamma rays from the
Galactic centre \citep[e.g.][among many others]{Bergstrom+98,
BaltzEdsjo99, GondoloSilk99, Morselli+02, Ullio+02, Stoehr+03, Prada+04,
Cesarini+04, Aharonian+06, Springel+08, CirelliPanci09, Fornasa+09,
BernalPalomares-Ruiz_10} as well as photons at other frequencies
\citep[e.g.][]{ColafrancescoMele01, Regis+08, RegisUllio08, Bergstrom+09,
Cholis+09, Pato+09, Profumo+10, Crocker+10}.
As the annihilation products travel through the surrounding medium, they heat
and ionize the gas,
potentially leaving an imprint on the cosmic microwave background
\citep{ChenKamionkowski04, Colafrancesco+04, PadmanabhanFinkbeiner05,
Mapelli+06, Zhang+06, Zhang+07,
Hooper+07, Cirelli+09, Galli+09, Cumberbatch+09, Kanzaki+10} and the HI 21~cm
spectral line \citep{Furlanetto+06, Vald+07, Chuzhoy08}. Heating
and ionization of the surrounding baryonic gas can also affect the formation of
the first stars \citep{Iocco+08, Spolyar+08, Natarajan+09,
Ripamonti+10} and the formation and evolution of galaxies \citep{Ascasibar07,
Ripamonti+07, Natarajan+08}.

\Referee
{
In this paper, we investigate the contribution of dark matter annihilation to
the total gas pressure and consider the possibility that it has a significant
effect on the rotation curve of spiral galaxies.
While rotation curves provided one of the first and most important pieces of
evidence for the existence of dark matter \citep[see e.g.][and references
therein]{SofueRubin01}, their shape in the inner regions of gas-rich dwarf and
low surface brightness (LSB) spiral galaxies is one of the outstanding issues in
modern cosmology \citep[see e.g.][for a recent discussion]{deBlok10}.

Observationally, rotation curves are found to rise approximately linearly with
radius, consistent with a constant density core in the dark matter distribution
\citep[e.g.][]{Flores&Primack94, Moore94, Burkert95, Kravtsov+98, Borriello+01,
deBlok+01, deBlokBosma02, Marchesini+02, Gentile+04, Donato+04_core, Spano+08,
Kuzio_de_Naray+08, deBlok+08, Oh+10} rather than the steep power law predicted
by cosmological N-body simulations \citep[e.g.][]{DubinskiCarlberg91,
Navarro+96, Navarro+97, Navarro+04, Navarro+10, FukushigeMakino+97,
FukushigeMakino01, Moore+98, Moore+99, Ghigna+98, Ghigna+00, JingSuto00,
Klypin+01, Power+03, Ascasibar+04, Fukushige+04, Hayashi+04, Reed+05,
Diemand+05, Diemand+08, AscasibarGottloeber08, Gao+08}.
Several modifications to the standard CDM scenario, such as warm
\citep{Colin+00, Sommer-Larsen+01},
repulsive \citep{Goodman00}, fluid \citep{Peebles00}, fuzzy \citep{Hu+00},
decaying \citep{Cen01}, annihilating \citep{Kaplinghat+00}, or self-interacting
\citep{Spergel+00, Yoshida+00, Dave+01} dark matter, and even alternative
theories of gravitation \citep[e.g.][]{McGaugh_deBlock98, SandersMcGaugh02,
Gentile+10} have been proposed in order to explain the discrepancy.

Here we focus on the energy density associated to electrons and positrons
arising from dark matter annihilations, neglecting other processes, such as dark
matter decay, or other annihilation products, such as protons
and antiprotons \citep[whose contribution is severely constrained by
recent observational data; see e.g.][]{Adriani+09_protons}.}
We adopt a model-independent
approach, in which all particles are created with the same initial
energy $E_0 \sim m_{\rm dm} c^2$. Results for a particular dark matter candidate
can be obtained by convolution with the appropriate source
function.

Since the characteristic energies involved are of the order of the mass of the
dark matter particle, and this mass is usually much larger than
the rest mass of the electron, electrons and positrons will be relativistic at
the moment of their creation. However, they can efficiently lose
their energy through different processes, such as inverse Compton scattering
(ICS), synchrotron radiation, Coulomb collisions, bremsstrahlung,
and ionization. Throughout this paper, we will often use the Lorentz factor
$\gamma$ to express the energy $E = \gamma m_{\rm e}c^2$, where
$m_{\rm e}$ denotes the rest mass of electron, and $c$ is the speed of light.

The pressure associated to these particles, hereafter referred to as ``dark
matter pressure'', is given by
\be
P_{\rm dm}(r) = \frac { m_{\rm e}
c^2 }{ 3 } \int^\infty_1 \frac{ {\rm d}n }{\rm d\gamma}(r,\gamma) \left( \frac{
\gamma^2 - 1 }{ \gamma } \right) {\rm d}\gamma,
\label{eqPressure}
\ee
where the electron-positron spectrum $\frac {{\rm
d}n}{\rm d\gamma}(r,\gamma)$ is the number density of particles with
Lorentz factor $\gamma$ at a radius $r$ from the centre of the dark matter halo.
The pressure gradient induces an acceleration
\be
a_{\rm dm}(r)
= - \frac{ 1 }{ \rho_{\rm g}(r) } \frac { {\rm d} P_{\rm dm}(r) }{ {\rm d}r },
\label{eqAcceleration}
\ee
where $\rho_{\rm g}(r)$ is the gas
density at radius $r$, that opposes the gravitational pull towards the centre,
affecting observable quantities such as the circular velocity
\be
v_{\rm c}(r) = \sqrt{ \frac{ G M(r) }{ r } + \frac{ r }{ \rho_{\rm g}(r) }
\deriv{P_{\rm dm}(r) }{ r } }. \label{eqVelocity}
\ee

It is our aim to show that, depending on the model parameters, the contribution
of dark matter pressure to the rotation curve may not be
negligible. Section~\ref{secModel} describes the procedure followed to estimate
the electron-positron spectrum. The ensuing dark matter pressure
is presented in Section~\ref{Results}, and the role of each astrophysical
parameter (gas density and ionization fraction of the interstellar
medium, intensity of the magnetic field, and dark matter density profile) is
discussed in detail. The effect on the rotation curve is
investigated in Section~\ref{ModelRotationCurves}, and our main conclusions are
briefly summarized in
Section~\ref{Conclusions}.

%--------------------------------------------------------------------------
 \section{The electron-positron spectrum}
 \label{secModel}
%--------------------------------------------------------------------------

%__________________________________
 \subsection{Propagation}
 \label{secProp}

The propagation of electrons and positrons through the interstellar medium (ISM)
is
determined by the diffusion-loss equation
\bea
\nonumber
\frac{ \partial }{ \partial t }
\frac{ {\rm d}n }{\rm d\gamma }(\textbf{x},\gamma)
&=&
\nabla
\left[
K(\textbf{x},\gamma) \nabla\frac{ {\rm d}n }{\rm d\gamma }(\textbf{x},\gamma)
\right]
\\
& + & \nonumber
\frac{ \partial }{ \partial\gamma }
\left[
b(\textbf{x},\gamma) \frac{ {\rm d}n }{\rm d\gamma }(\textbf{x},\gamma)
\right]
\\
& + & Q(\textbf{x},\gamma).
\eea
We assume a diffusion coefficient of the form
\be
K(\gamma) = K_0 \gamma^\delta
\ee
with $K_0 =
1.67\times10^{25} \rm{cm^2s^{-1}}$ and $\delta=0.7$, independent of Galactic
location \citep[MED model in][]{Donato+04}. The energy loss rate
\be
b(\textbf{x},\gamma) \equiv -\frac{ {\rm d}\gamma }{ {\rm d}t }(\textbf{x},
\gamma) = \sum_i b_i(\textbf{x}, \gamma)
\ee
is a sum over the
relevant physical processes, and the source term $Q(\textbf{x},\gamma)$
represents the instantaneous electron-positron injection rate.

Given enough time, the electron-positron population will approach
a steady-state distribution, $\frac{\partial}{\partial
t}\frac{{\rm d}n}{\rm d\gamma}(\textbf{x},\gamma)=0$. Assuming
that $b(\textbf{x}, \gamma)$ varies smoothly in space, the
particle spectrum fulfills the relation
\be
\frac{ \partial
y(\textbf{x},\gamma) }{ \partial\gamma } + \frac{ K(\gamma) }{
b(\gamma) } \nabla^2 y(\textbf{x}, \gamma) = -Q(\textbf{x},
\gamma), \ee where \be y(\textbf{x},\gamma) \equiv b(\gamma)
\frac{ {\rm d}n }{\rm d\gamma }(\textbf{x},\gamma).
\ee
Imposing
$\frac{{\rm d}n}{\rm d\gamma}(\textbf{x},\gamma)=0$ at infinity,
one obtains the Green's function
\be
G(\textbf{x}, \gamma,
\textbf{x}_{\rm s}, \gamma_{\rm s}) = \frac { \exp\left(-\frac
{\mid\textbf{x}-\textbf{x}_s\mid ^2} {2\Delta\lambda^2} \right) }
{ \left(2\pi\Delta\lambda^2\right)^{3/2} } \
\Theta(\gamma-\gamma_{\rm s})
\ee
and either the image charges
method or an expansion over the eigenfunctions of the linear
differential operator may be used to derive the Green's function
for other boundary conditions \citep[see
e.g.][]{BaltzEdsjo99,Delahaye+09}. The electron-positron spectrum
is given by

\be
\frac{{\rm d}n}{\rm d\gamma}(\textbf{x},\gamma)
\!=
\!\frac{ 1 }{ b(\textbf{x},\gamma) }
\!\int^\infty_\gamma\!\!\!\!\!\!\!\rm d\gamma_{\rm s}
\!\int_0^\infty\!\!\!\!\!\!\!\rm d^3\!\!\textbf{x}_s
\frac
{
 \exp
 \left(
   \!-\frac{\mid\textbf{x}-\textbf{x}_s\mid^2}{2\Delta\lambda^2}
  \right)
}
{ \left(2\pi\Delta\lambda^2\right)^{3/2} }
\textit{Q}(\textbf{x}_{\rm s}, \gamma_{\rm s})
\ee
where the quantity
\be
 \Delta\lambda^2 = \lambda^2(\gamma)-\lambda^2(\gamma_{\rm s})
\ee
is related to the characteristic diffusion length of the electrons and
positrons, and $\gamma_{\rm s}$ denotes their initial energy.
The variable $\lambda$ is defined as
\be
\lambda^2(\gamma) = \int_\gamma^\infty \frac{ 2 K(\gamma) }{ b(\gamma) }\rm
d\gamma.
\ee
Considering the dark matter halo as a spherically-symmetric source, the spatial
integral can be reduced to one dimension, and the
electron-positron spectrum is finally given by the expression
\bea
\nonumber
\frac{ {\rm d}n }{\rm d\gamma }(r,\gamma) \!\!\!\!&=&\!\!\!\!
\frac{ 1 }{ b(\gamma) } \frac{ \exp\left( - \frac{ r^2 }{ 2\Delta\lambda^2 }
\right)}
     { \left( 2\pi r^2 \Delta\lambda^2 \right)^{1/2} }
\\
&\times&\!\!\!\!\!\! \nonumber
\Big\{
\, \int^\infty_\gamma\!\!\!\! {\rm d} \gamma_{\rm s}
\int_0^\infty\!\!\!\! {\rm d} r_{\rm s}
\ r_{\rm s}
\ \exp\left( -\frac{ r_{\rm s}^2 }{ 2 \Delta\lambda^2 } \right)
\\
& &
\left[
\exp\left( \frac{ r r_{\rm s} }{ \Delta\lambda^2 } \right)
-
\exp\left( - \frac{ r r_{\rm s} }{ \Delta\lambda^2 } \right)
\right]
Q(r_{\rm s}, \gamma_s)\Big\}
\label{eqSpectrum}
\eea

%__________________________________
 \subsection{Loss rates}
 \label{secLoss}

Electrons and positrons can lose their energy by several physical processes as
they move through the ISM. We consider ICS of cosmic microwave
background (CMB) and starlight photons, synchrotron radiation, Coulomb
collisions, bremsstrahlung, and ionization of neutral hydrogen atoms.

The energy loss rates depend on the energy of the particle.
High-energy electrons and positrons mainly lose energy by ICS
\citep[e.g.][]{Sarazin99}.
The relevant loss function is
\be
b_{\rm ICS}(\gamma) =
\frac{ 4 }{ 3 } \frac{ \sigma_{\rm T} }{ m_{\rm e} c } \gamma^2 U_{\rm rad}
\label{eqICloss}
\ee
where $\sigma_{\rm T}$ is the Thomson cross section and
\begin{equation}
 U_{\rm rad} = U_{\rm CMB} + U_{\rm stars} + U_{\rm dust} \approx 0.9\ {\rm
eV~cm^{-3}}
\end{equation}
is the combined radiation energy density of the CMB, starlight, and thermal dust
emission \citep[see e.g.][]{PorterMoskalenko05}.

Synchrotron radiation is another important loss mechanism at high energies.
The expression for the loss rate is similar to that of ICS, substituting the
radiation energy density in equation~(\ref{eqICloss}) by the energy density of
the magnetic field, $U_{\rm B}=B^2/(8\pi)$, where $B$ is the intensity of the
magnetic field:
\be
b_{\rm syn}(\gamma) =
\frac{ 4 }{ 3 } \frac{ \sigma_{\rm T} }{ m_{\rm e} c } \gamma^2 U_{\rm B}.
\label{eqsynloss}
\ee
% Both equations (\ref {eqICloss}) and (\ref {eqsynloss}) assume $\gamma\gg 1$.

For lower-energy electrons and positrons, Coulomb interactions with the thermal
plasma must be taken into account.
The loss rate is approximately \citep{Rephaeli79}
\be
b_{\rm Coul}(\gamma) \approx
1.2 \times 10^{-12} n_{\rm e}
\left[ 1 + \frac{ \ln( \gamma / n_{\rm e} ) }{ 75 } \right]\ \ \rm{s^{-1}}
\ee
where $n_{\rm e}$ is the number density of thermal electrons.

Collisions with thermal ions and electrons also produce radiation through
bremsstrahlung.
The loss rate due to bremsstrahlung can be approximated as
\citep{BlumenthalGould70}
\be
b_{\rm brem}(\gamma) \approx 1.51 \times 10^{-16} n_{\rm e} \gamma \left[
\ln(\gamma) + 0.36 \right]\ \ \rm{s^{-1}}.
\ee

Additional energy losses come from the ionization of hydrogen atoms. The loss
rate is given in \citet{Longair81},
\bea
\nonumber b_{\rm
ion}(\gamma) \!\!\!\!&=&\!\!\!\! \frac{ q_{\rm e}^4 n_{\rm H} }
     { 8\pi \epsilon_0^2 m_{\rm e}^2 c^3 \sqrt{ 1 - \frac{1}{\gamma^2} } }
\times
\Big[~
\ln \frac{ \gamma ( \gamma^2 - 1 ) }{ 2 \left( \frac{ I }{ m_{\rm e} c^2 }
\right)^2 }\\
& & \!\!\!\! - \left( \frac{ 2 }{ \gamma } - \frac{ 1 }{ \gamma^2 } \right) \ln2
+ \frac{ 1 }{ \gamma^2 } + \frac{ 1 }{ 8 } \left( 1 - \frac{ 1
}{ \gamma } \right)^2 ~\Big]
\eea
where $n_{\rm H}$ is the number density of
hydrogen atoms, $q_{\rm e}$ is the charge of electron, $\epsilon_0$
is the permittivity of free space and $I$ is the ionization energy of the
hydrogen atom. The number density of thermal electrons and neutral
atoms can be expressed in terms of the total ISM gas density $\rho_{\rm g}$ and
the ionization fraction $X_{\rm ion}$ as
\be
n_e = \frac{
\rho_{\rm g} }{ m_{\rm p} } X_{\rm ion} \ee and \be n_{\rm H} = \frac{ \rho_{\rm
g} }{ m_{\rm p} } ( 1 - X_{\rm ion} )
\ee
respectively.

%__________________________________
 \subsection{Source term}

Since the electrons and positrons in our model originate from the annihilation
of dark matter particles, the production rate is dictated by the dark matter
number density and
the annihilation rate into electron-positron pairs,
\be
Q(r,\gamma) =
n_{\rm dm}(r)\ n_{\rm dm^*}(r)
\ \langle \sigma v \rangle_{e^\pm}
\ \deriv{N_{e^\pm}}{\gamma}
\ee
where $n_{\rm dm}$ and $n_{\rm dm^*}$ denote the number densities of dark matter
particles and anti-particles, respectively, $\langle\sigma v\rangle_{e^\pm}$ is
the thermal average of the annihilation cross-section times the dark matter
relative velocity, and $\deriv{N_{e^\pm}}{\gamma}$ is the injection spectrum of
electrons and positrons in the final state.

Assuming all electrons and positrons are injected with the same energy $\gamma_0
\sim m_{\rm dm} / m_{\rm e}$, 
\be 
Q(r,\gamma) = Q_0(r)\
\delta(\gamma-\gamma_0) 
\ee 
where 
\be Q_0(r) = 2\, \left[ \frac{ \rho_{\rm
dm}(r) }{ m_{\rm dm} } \right]^2 \langle \sigma v \rangle_{e^\pm}
\label{eqQ0} 
\ee 
\Referee{is the local production rate per unit volume per unit time}
and $\delta(\gamma-\gamma_0)$ denotes a Dirac delta function.

Although this is a rather coarse
approximation, it has the advantage of being absolutely
model-independent. Moreover, the contribution of electrons and positrons to the
gas pressure will be mostly determined by their total number and
average initial energy, with the details of the injection spectrum playing only
a minor role. The factor of 2 in equation~(\ref{eqQ0}) accounts
for one electron and one positron produced per annihilation event, and
self-conjugate (Majorana) dark matter particles have been assumed. If dark matter
particles and anti-particles were different, $n_{\rm dm} = n_{\rm dm^*} =
\rho_{\rm dm}(r)/(2m_{\rm dm})$ and $Q_0$ would decrease by a factor
of four. For the dark matter density $\rho_{\rm dm}(r)$, we consider a perfectly
spherically-symmetric halo described by a density profile of
the form 
\be 
\rho_{\rm {dm}}(r) = \frac { \rho_{\rm s} } {
\left( \frac { r }{ r_{\rm s} } \right)^\alpha
\left( 1 + \frac { r }{ r_{\rm s} } \right)^{3-\alpha}
},
\label{eqDMProfile1}
\ee
where $r_{\rm s}$ and $\rho_{\rm s}$
denote a characteristic density and radius of the halo,
respectively, and the $\alpha$ is the inner logarithmic slope of
the density profile. Local inhomogeneities that would boost the
expected signal, such as small-scale clumpiness or the presence of
subhaloes, are not taken into account.
\Referee
{
The shape of the dark matter density profile in the inner regions is far from
being a settled question.
As stated in the introduction, N-body simulations suggest that, at least in the
absence of baryons, the profile should be quite steep near the centre ($\alpha
\sim 1$), in apparent contradiction with observations.
Traditionally, it has been argued that the presence of gas and stars makes the
profile even steeper due to the effects of adiabatic contraction
\citep{Blumenthal+86}, although some recent claims have also been made in the
opposite direction \citep[e.g.][]{El-Zant+01, Mashchenko+06, Oh+_10}.
Given the current uncertainties, we have left the inner slope of the density
profile as a free parameter of the model.
}

%__________________________________
\begin{figure}
\centering \includegraphics[width=8cm]{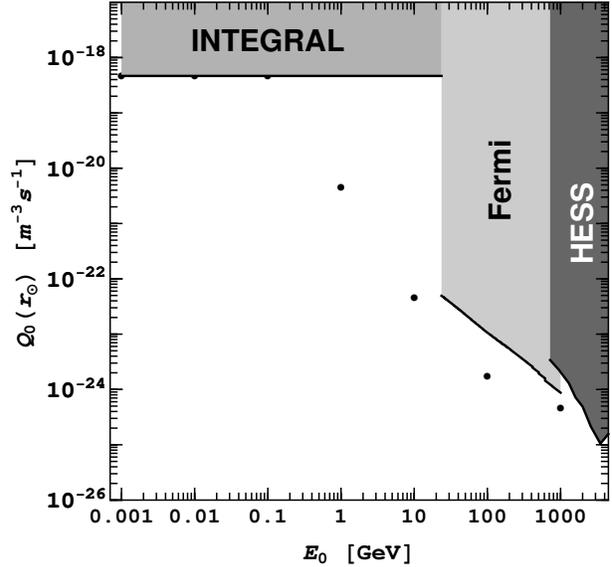}
\caption{Exclusion regions and production rates at the position of the Sun:
black dots are the adopted values of the instantaneous production rate $Q_0$
(see Table~\ref{tabProductionrate}). Shaded regions above the black
lines are excluded by INTEGRAL, Fermi and HESS data.} \label{figQ0}
\end{figure}
%__________________________________

%__________________________________
\begin{table}
\begin{center}
\begin{tabular}{clcc}
$\gamma_0$ & ~~$~~\frac{ E_0 }{\rm GeV }$ & $\frac{ \langle \sigma v
\rangle_{e^\pm}}{ \rm cm^3\ s^{-1} }$ & $\frac{ Q_0(r_\odot)}{\rm
m^{-3}~s^{-1}}$\\ \hline
$2 \times 10^0$ & $1.022 \times 10^{-3}$ & $2.6 \times 10^{-30}$ & $4.7\times
10^{-19}$\\
$2 \times 10^1$ & $1.022 \times 10^{-2}$ & $2.6 \times 10^{-28}$ & $4.7\times
10^{-19}$\\
$2 \times 10^2$ & $1.022 \times 10^{-1}$ & $2.6 \times 10^{-26}$ & $4.7\times
10^{-19}$\\
$2 \times 10^3$ & $1.022 \times 10^0   $ & $2.6 \times 10^{-26}$ & $4.7\times
10^{-21}$\\
$2 \times 10^4$ & $1.022 \times 10^1   $ & $2.6 \times 10^{-26}$ & $4.7\times
10^{-23}$\\
$2 \times 10^5$ & $1.022 \times 10^2   $ & $1.0 \times 10^{-25}$ & $1.8\times
10^{-24}$\\
$2 \times 10^6$ & $1.022 \times 10^3   $ & $2.6 \times 10^{-24}$ & $4.7\times
10^{-25}$
\end{tabular}
\end{center}
\caption{Initial Lorentz factors, energies, \Referee{cross-sections} and
production rates at the position
of the Sun used in our
calculations.}
\label{tabProductionrate}
\end{table}
%__________________________________

\Referee{For similar reasons, we also consider} the injection energy as a free
parameter and investigate
values of the initial Lorentz factor $\gamma_0$ between $2$ and
$2\times 10^6$, corresponding to energies $E_0 = \gamma_0 m_{\rm e} c^2$ from
about 1~MeV to 1~TeV.
\Referee
{
The production rate $Q_0$, on the other hand, is
strongly constrained by different Galactic observations.
At high energies, we consider observations of the
electron-positron spectrum in the solar neighbourhood by HESS and the Large
Area Telescope (LAT) on board the Fermi satellite \citep{Aharonian+08,Abdo+09}.
More specifically, the predicted amount of electrons and positrons cannot exceed
the observed values for \emph{any} Lorentz factor $\gamma$.
Given the energy dependence of the observed spectrum, $\left[ \deriv{n}{E}
\right]_{\rm obs} \sim E^{-3}$, and the energy losses, $b(E)\sim E^2$, the most
restrictive constraint comes from the spectrum near the injection energy, where
propagation can be safely neglected and  $\left[ \deriv{n}{E} \right]_{\rm
model} \approx \frac{Q_0}{b} \propto E^{-2}$.
The maximum production rate allowed by the data can then be expressed as
\be
Q_0(r_\odot) < b(\gamma_0)\, \left[ \deriv{n}{E} \right]_{\rm
obs}\!\!\!\!\!\!\!\!(\gamma_0)
\ee 
Another, completely independent upper limit, valid at all energies, can
be obtained from the observed intensity of the 511 keV line that measures the
positron annihilation rate at the Galactic centre.
In order to fully explain the line with dark matter annihilations, it is
necessary that \citep{Ascasibar+06}
\be
\frac{ \langle \sigma v \rangle_{511} }{ 2.6 \times 10^{-30}\ {\rm cm^3\ s^{-1}}
} = \left( \frac{ m_{\rm dm} c^2 }{\rm 1~MeV } \right)^2 
\ee
so one just have
\be
Q_0(r_\odot) < 2\, \left[ \frac{ \rho_{\rm dm}(r_\odot) }{ m_{\rm dm} }
\right]^2 \langle \sigma v \rangle_{511}
\label{eqProductionRate3}
\ee
in order not to overproduce the observed signal.

The corresponding exclusion regions are shown in Figure~\ref{figQ0}, together
with the production rates used in our calculation at the position of the Sun,
$r_\odot=8.5~\rm{kpc}$.
These conditions constitute strict upper limits, since astrophysical sources
will also contribute to the relativistic particle budget, but the annihilation
cross-sections they imply are comparable to or larger than the ones required to
explain the cosmic dark matter density,
\be
\Omega_{\rm dm} \sim \frac{ 10^{-26}\ {\rm cm^3\ s^{-1}} }{ \langle \sigma v
\rangle_{e^\pm} }
\ee
}
\Referee{
Therefore, we have set the injection rate according to
equation~(\ref{eqProductionRate3}) for $m_{\rm dm} c^2 \le 100$~MeV, while a
cross-section compatible with the relic density constraint, $\langle \sigma v
\rangle_{e^\pm} = 2.6 \times 10^{-26}\ {\rm cm^3\ s^{-1}}$, has been assumed for
$100$~Mev$\le m_{\rm dm} c^2 \le 10$~GeV, and slightly larger values
\citep[based on the positron excess observed by PAMELA;][]{Adriani+09_positrons}
have been used for $m_{\rm dm} c^2 \approx 100$~GeV and 1~TeV.
Numeric values are given in Table~\ref{tabProductionrate}.
}

%--------------------------------------------------------------------------
 \section{Dark matter pressure}
 \label{Results}
%--------------------------------------------------------------------------

Apart from the initial energy and injection rate of the electron-positron pairs,
related to the nature of the dark matter particle, there are
many astrophysical parameters that determine the contribution of dark matter
annihilation to the total gas pressure. We will first define a
canonical model based on observations of the Milky Way and then investigate the
effect of each individual component by varying the values of the
adopted parameters one by one. In all cases, we calculate the electron-positron
spectrum as described in the previous section, and then estimate
the dark matter pressure according to expression~(\ref{eqPressure}).

%__________________________________
 \subsection{Canonical Milky Way model}

Our canonical model assumes a dark matter density profile with $\alpha=1$
\citep{Navarro+97}, $r_{\rm s}=17$~kpc and $\rho_{\rm s}=6 \times
10^{-22}$~kg~m$^{-3}$, consistent with dynamical models of the Milky Way
\citep[e.g.][]{DehnenBinney98, Klypin+02}. The virial mass of the
Galaxy is thus $10^{12}$~M$_\odot$, and the local dark matter density is
$\rho_{\rm dm}(r_\odot)\,c^2 = 0.3$~GeV~cm$^{-3}$. The ISM is mainly
composed by neutral hydrogen atoms ($X_{\rm ion} = 0$) with number density
$\rho_{\rm g}/m_{\rm p} \sim 1$~cm$^{-3}$
\citep{DehnenBinney98,Ferriere01,Robin+03}, and it is permeated by a uniform,
tangled magnetic field whose intensity is $B \sim 6~\mu$G
throughout the Galaxy \citep{Ferriere01,Beck01,AscasibarDiaz10}.

%__________________________________
\begin{figure}
\begin{center}
\includegraphics[width=8cm]{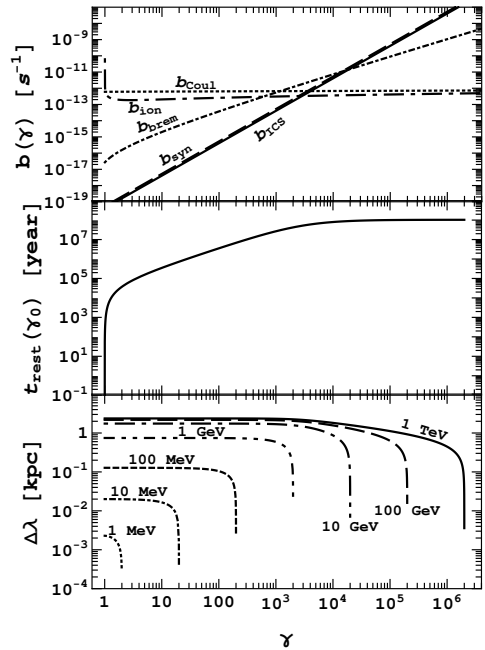}
\end{center}
\caption
{\Referee
{Energy losses, equilibrium time scales, and characteristic diffusion lengths of
electrons and positrons for $\rho_{\rm g}/m_{\rm p} =
1~{\rm cm^{-3}}$, $B=6\ \rm {\mu G}$, and $X_{\rm ion}=0.5$.
On the top panel, Coulomb collisions, ionization, bremsstrahlung, synchrotron
radiation, and
inverse Compton scattering are represented by dotted, dot-long
dashed, dot-short dashed, dashed, and solid lines, respectively.}
}
\label{figLoss}
\end{figure}
%__________________________________

The contribution of the individual loss terms described in Section~\ref{secLoss}
is plotted on the top panel in Figure~\ref{figLoss} for a
model similar to the canonical one, but with $X_{\rm {ion}}=0.5$ (in order to
have a non-zero contribution from Coulomb collisions and
bremsstrahlung). ICS and synchrotron emission, being proportional to $\gamma^2$,
dominate at high energies, $\gamma>10^4$. Bremsstrahlung is
important in the intermediate range $10^3 < \gamma < 10^4$, and Coulomb
collisions and ionization, roughly independent on $\gamma$, dominate at
low energies, $\gamma<10^3$.
\Referee
{
The time taken by the electron-positron population to reach equilibrium is of
the same order as the time
\be 
t_{\rm rest}(\gamma_0)=\int^{\gamma_0}_1 \frac{1}{b(\gamma)}\ \dd \gamma, 
\ee
that the particles take to loose all their energy, shown on the middle panel in
Figure~\ref{figLoss} as a function of $\gamma_0$.
Although this time may be larger than the orbital time at $r \sim 500$~pc ($T
\sim 10$~Myr, assuming $v \sim 220$~km~s$^{-1}$) for $m_{\rm dm} > 100$~MeV, a
steady state will be reached as long as the conditions (dark matter and gas
densities, magnetic field, etc.) evolve on timescales longer than $\sim
100$~Myr, which is relatively short in astrophysical terms.
}

%__________________________________
\begin{figure*}
\includegraphics[width=8cm]{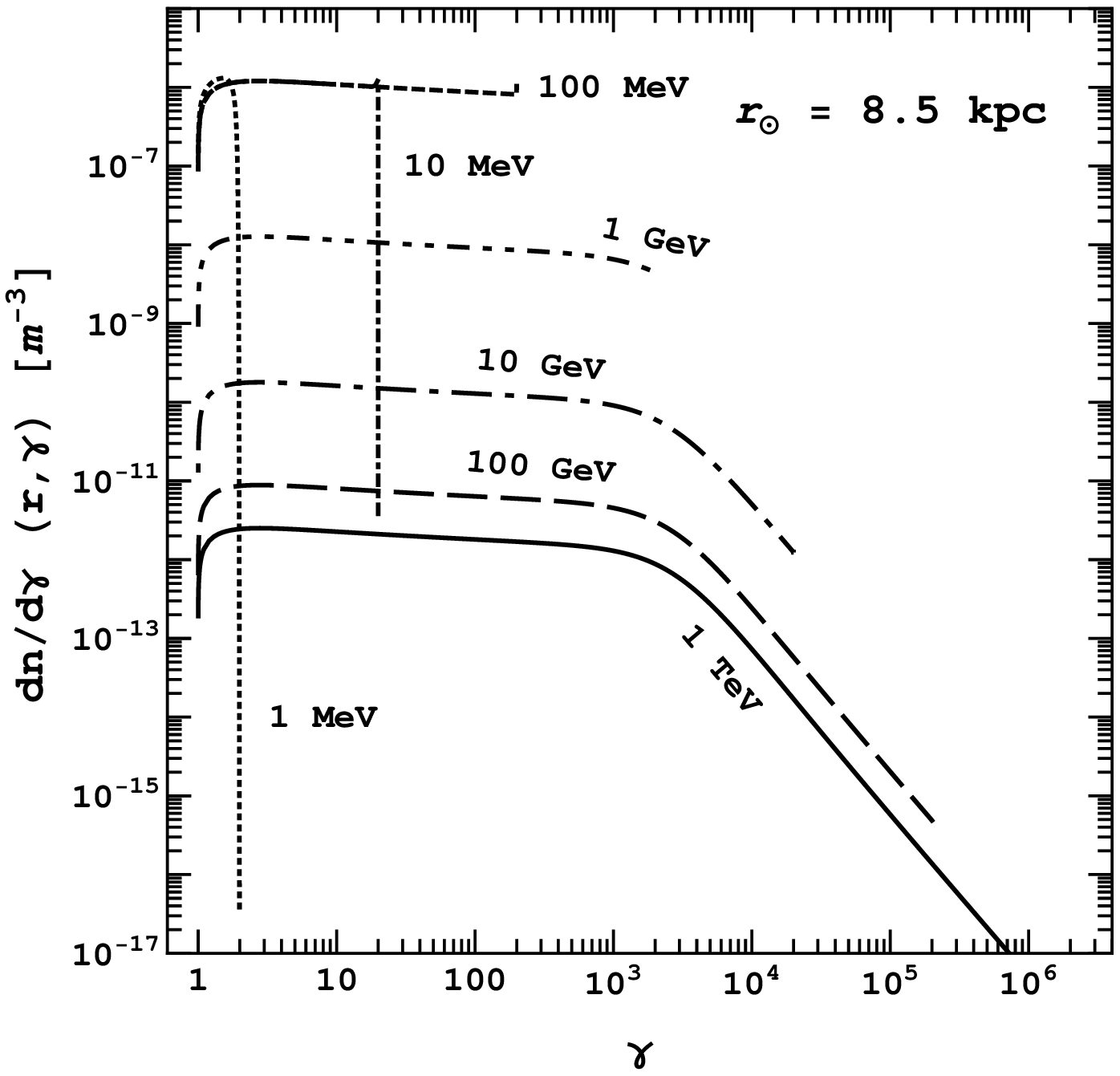}
\hspace{1cm}
\includegraphics[width=8cm]{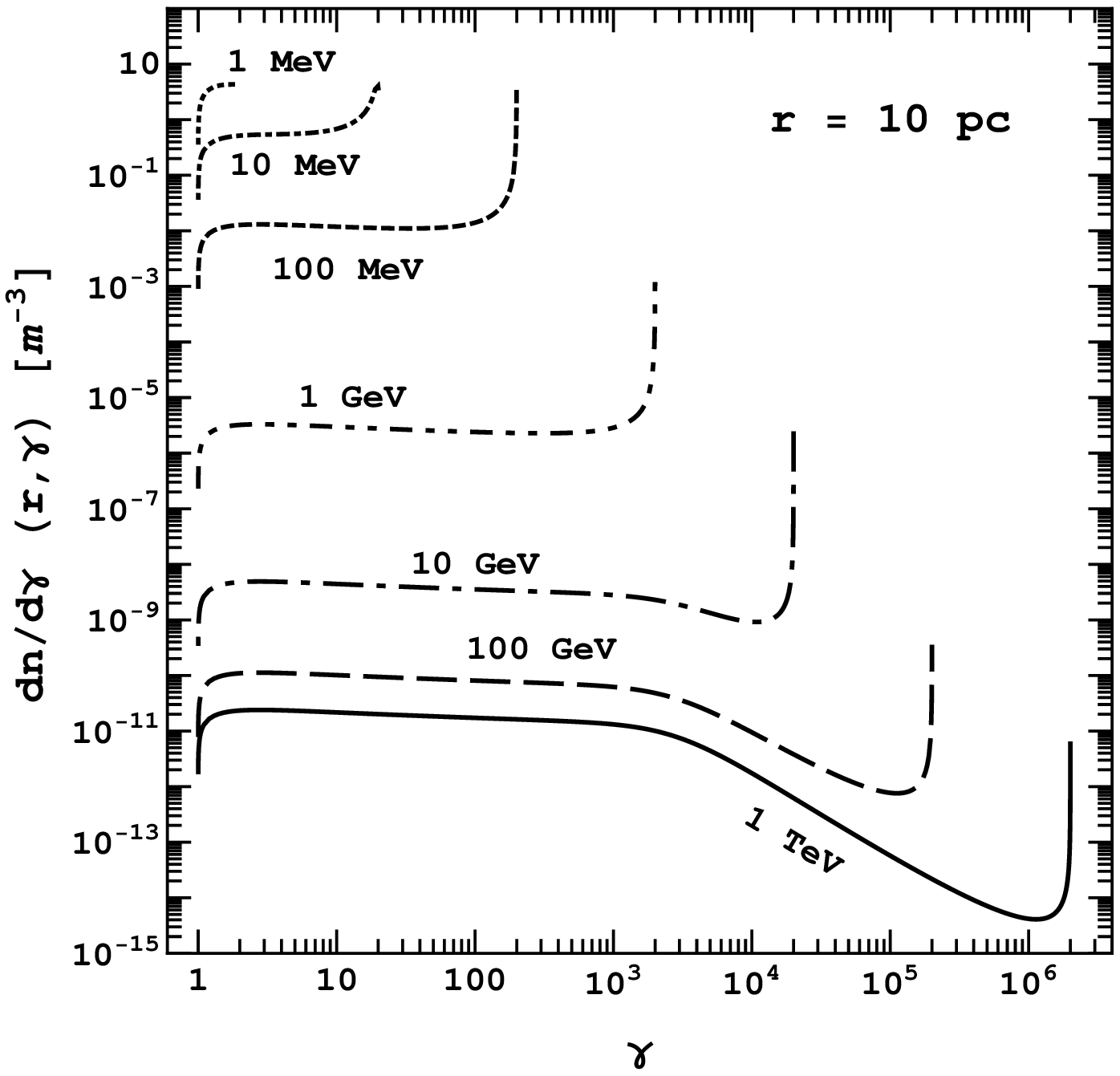}
\caption {Electron-positron spectrum at the position of the Sun (left) and at
$r=10$~pc (right) for our canonical model of the Milky Way and
different values of the injection energy $E_0$. } \label{figCanonicalSpectrum}
\end{figure*}
%__________________________________
%__________________________________
\begin{figure*}
\includegraphics[width=8cm]{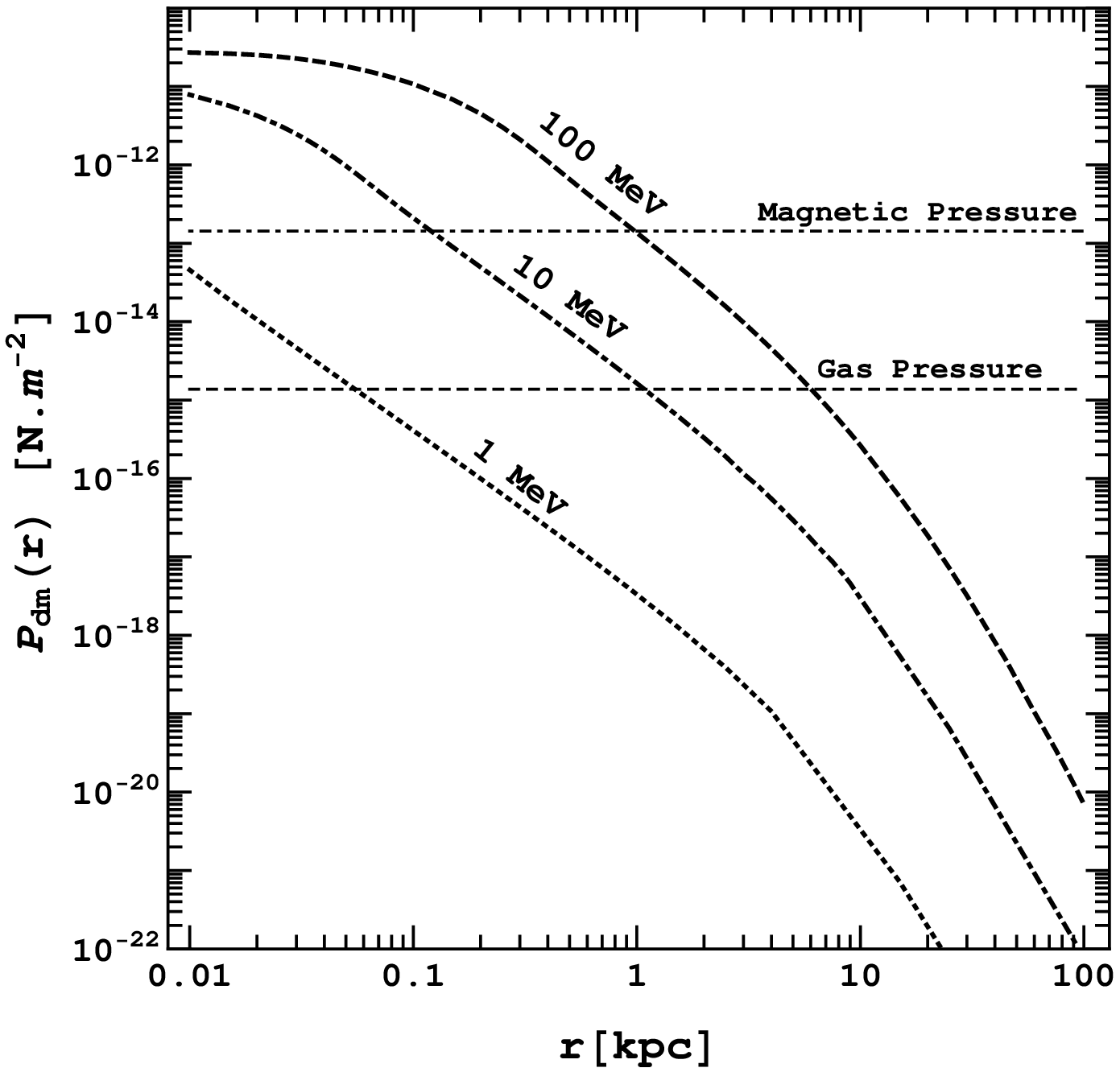}
\hspace{1cm}
\includegraphics[width=8cm]{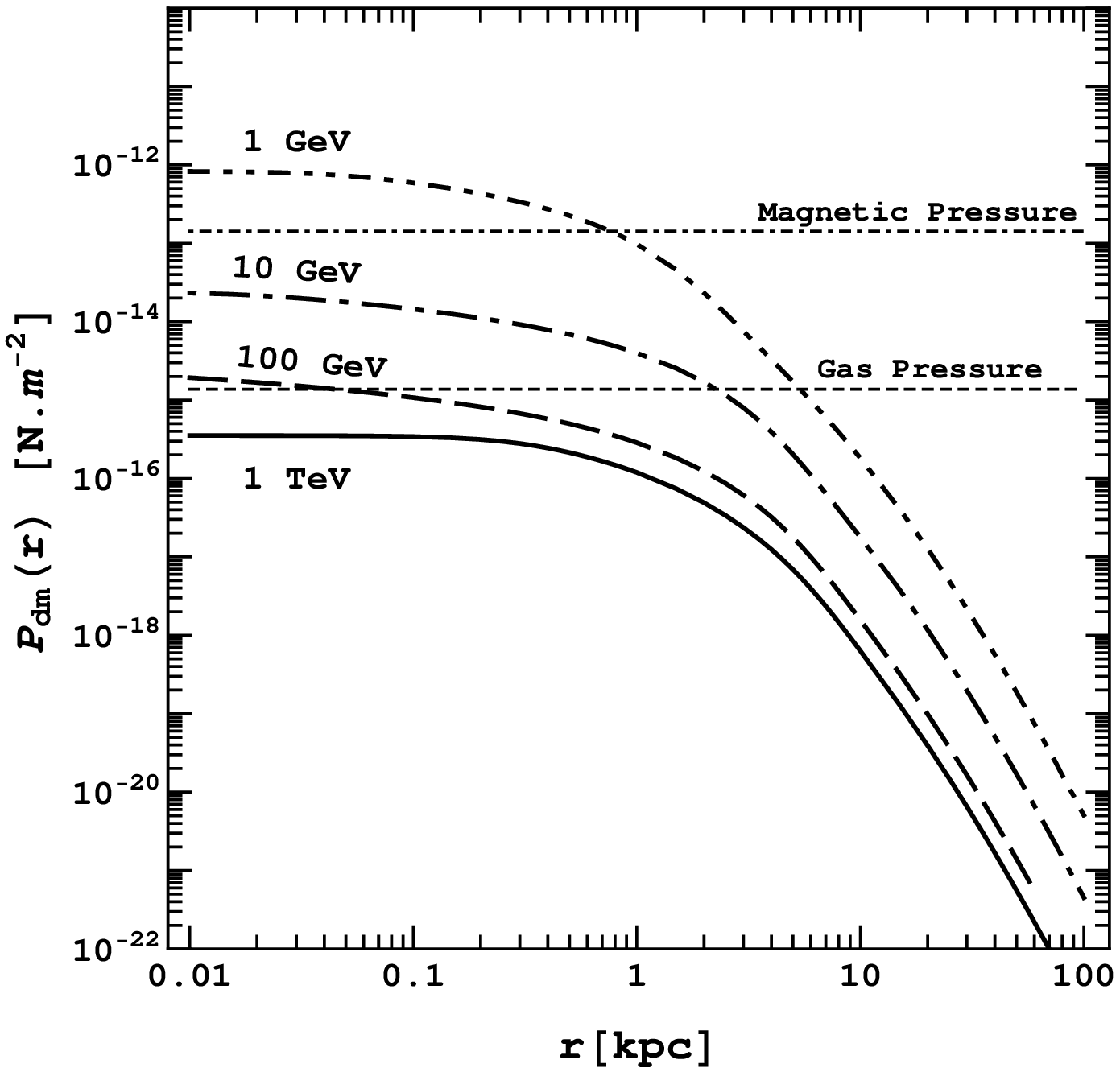}
\caption{\Referee{Dark matter pressure as a function of radius in our canonical
model
compared with gas and magnetic pressure.}}
\label{figCanonicalPressure}
\end{figure*}
%__________________________________

The steady-state electron-positron spectrum at the position of the Sun,
$r_\odot=8.5\
\rm{kpc}$, is shown on the left panel in Figure \ref{figCanonicalSpectrum} for
different values of the initial energy $E_0$.
As stated above, all of the electrons and positrons are generated with the same
$\gamma_0$, according to Table~\ref{tabProductionrate}.
Propagation through the Galaxy and energy losses are accounted for by
equation~(\ref{eqSpectrum}).
The shape of the resulting spectrum is determined by the value of $\gamma_0$,
the production rate $Q_0(r)$, the loss rates $b(\gamma)$ implied by the values
of $\rho_{\rm g}$, $B$, and $X_{\rm ion}$, and the diffusion coefficient
$K(\gamma)$.
Nevertheless, some insight may be gained by neglecting diffusion.
For $K_0 = 0$,
\be
\deriv{ n }{ \gamma }(r_\odot,\gamma) = \frac{ Q_0(r_\odot) }{ b(\gamma) };
\ee
the electron-positron spectrum is almost flat when ionization dominates the
energy losses, and there is a transition at $\gamma\sim10^3$ ($E \sim 1$~GeV) to
the ICS-synchrotron regime, where $\deriv{ n }{ \gamma } \propto \gamma^{-2}$.
For low injection energies (1 MeV to 100 MeV), the normalizations of the spectra
are identical because the value of $Q_0$ is only constrained by the INTEGRAL
data, whereas other constraints impose lower values at higher energies (see
Figure~\ref{figQ0}).
In all cases, the spectra are cut at the injection energy $\gamma_0$ since no
acceleration mechanism is included in our model.

% %__________________________________
% \begin{figure}
% \centering
% \includegraphics[width=8cm]{figs/DiffusionLength.eps}
% \caption{Characteristic diffusion lengths of the electrons and positrons for
% different values of the injection energy $E_0$.}
% \label{figDiffusion}
% \end{figure}
% %__________________________________

The electron-positron spectrum closer to the centre of the Galaxy ($r=10$~pc) is
shown on the right panel in Figure \ref{figCanonicalSpectrum}.
In general terms, the overall normalizations are higher than at the position of
the Sun because of the higher dark matter density, and there is
a sharp spectral feature near $E_0$. The characteristic diffusion scale
$\Delta\lambda$ plays an important role in both cases. As can be seen on
the bottom panel of Figure~\ref{figLoss}, $\Delta\lambda$ depends on the Lorentz
factor of the
electrons and positrons. It is zero at the injection value, and
it rapidly increases for lower energies until it saturates at a maximum value
that depends on $E_0$. The spectrum at a given $\gamma$ probes the
effective value of the production rate $Q(r)$, averaged over the diffusion
scale. This is not very relevant at the position of the Sun because
the dark matter density does not vary much on kpc scales, but it becomes more
important as one moves towards the central density cusp. For
$\gamma\ll\gamma_0$, $\Delta\lambda$, and thus the effective production rate, is
independent on $\gamma$. The larger $E_0$, the larger the
smoothing scale, and therefore the smaller the average density and the contrast
with respect to the normalization at 8.5~kpc. As long as
$\Delta\lambda$ is constant, the shape of the spectrum remains the same, flat
for low Lorentz factors and proportional to $\gamma^{-2}$ in the
inverse Compton regime. Near the injection energy, $\Delta\lambda$ becomes very
small, the effective production rate approaches the local source
term $Q_0(r)$, much higher than the smoothed value, and the spectrum rises
steeply just before the cutoff.

%__________________________________
\begin{figure*}
\includegraphics[width=5.5cm]{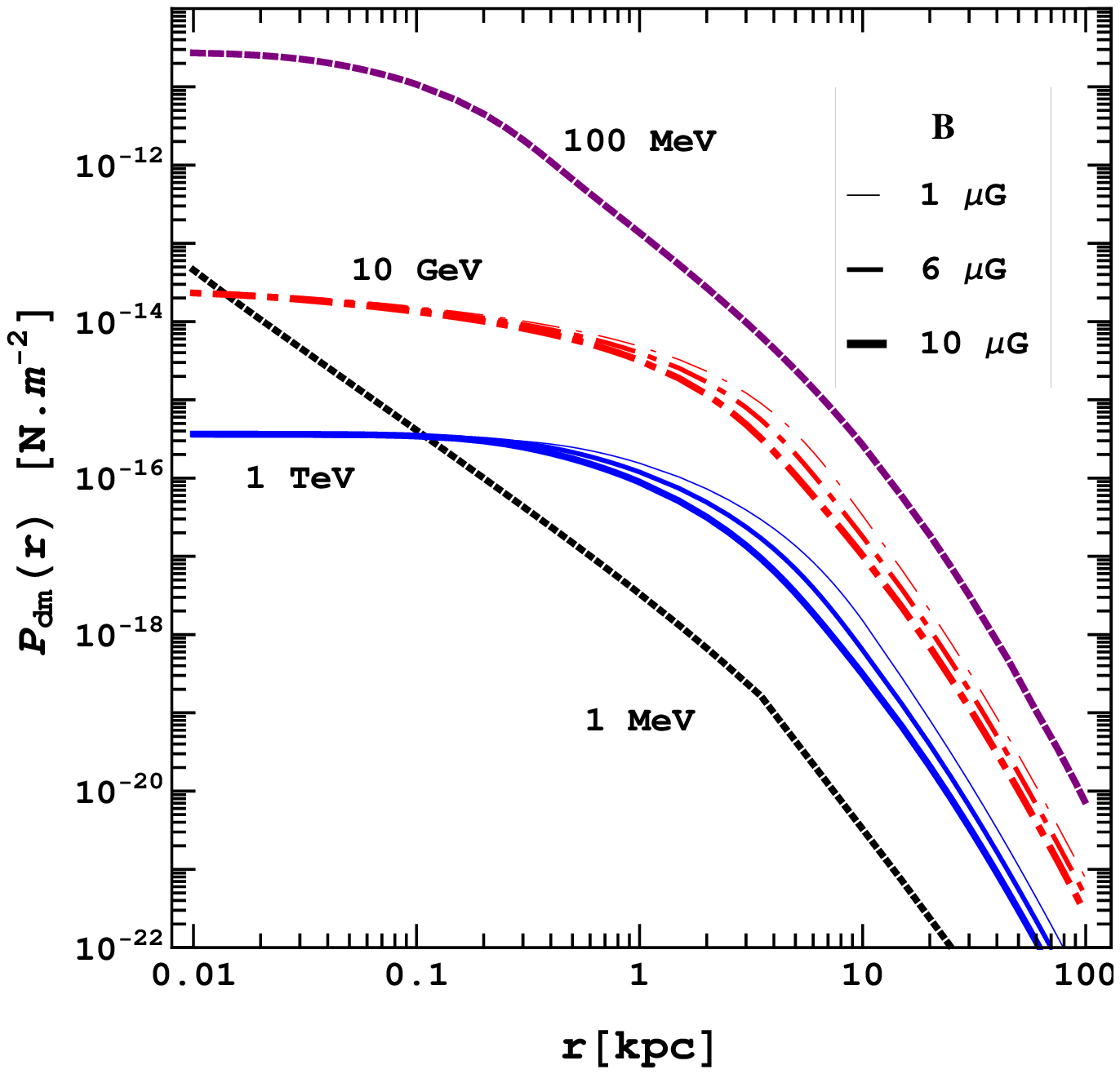}
\hfill\includegraphics[width=5.5cm]{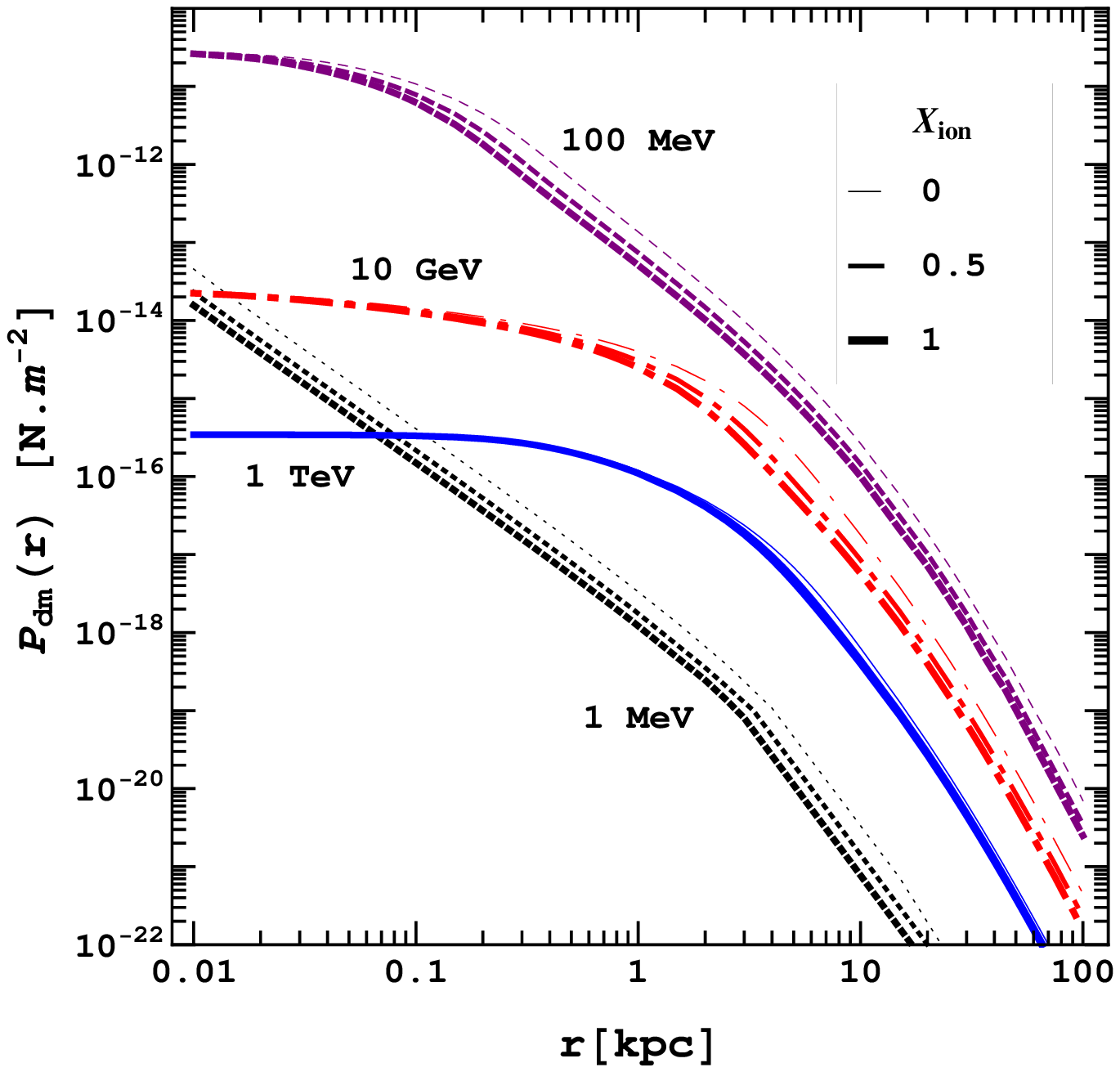}
\hfill\includegraphics[width=5.5cm]{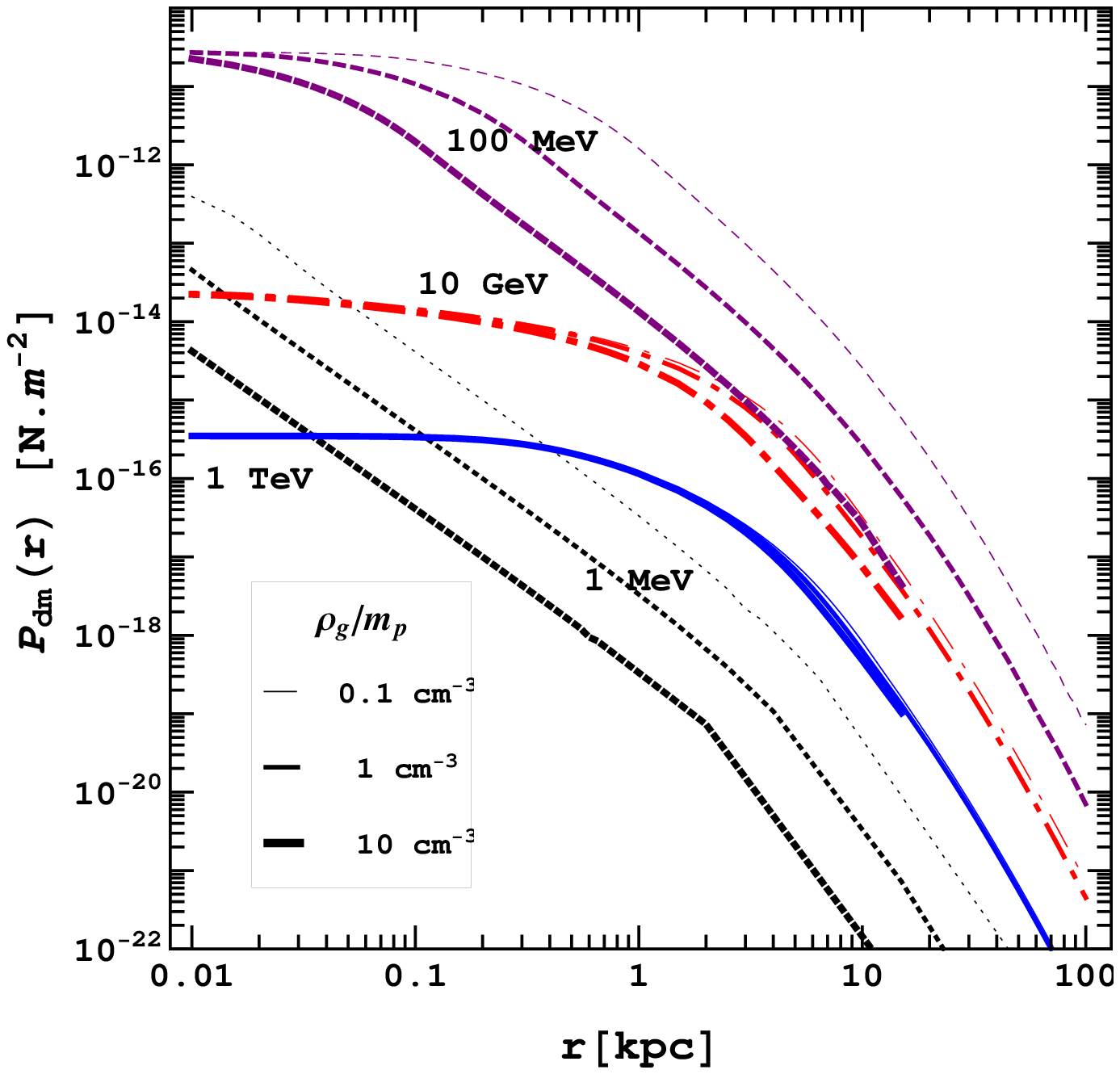} \caption{Effect
of varying the magnetic field intensity from 1 to $10\ \mu\rm G$ (left), the
ionization fraction $X_{\rm ion}$ from 0 to 1 (centre), and the ISM
gas density from $\rho_{\rm g}/m_{\rm p} =0.1$ to 10~cm$^{-3}$ (right). The
pressure profiles for $E_0=1$~MeV, 100~MeV, 10~GeV, and 1~TeV are
represented by dotted, dashed, dot-dashed, and solid lines, respectively.}
\label{figAstro}
\end{figure*}
%__________________________________

Finally, the contribution of dark matter annihilations to the gas pressure at a
given radius can be obtained by substituting the
electron-positron spectrum in equation~(\ref{eqPressure}). Results for different
values of $\gamma_0$ compared with gas and magnetic pressure in
the galaxy are presented in Figure~\ref{figCanonicalPressure},
\Referee
{
compared to the thermal pressure of the gas $P_{\rm gas} = nkT$ (where
$n=\rho_{\rm g}/m_{\rm p}=1$~cm$^{-3}$ is the gas density, $k$ is the Boltzmann
constant, and we have assumed a temperature $T=100\ \rm K$, appropriate for the
neutral gas in the Galactic disk) and the magnetic pressure $P_{\rm mag} =
B^2/{8\pi}$, with $B=6~\mu$G.
}

For low injection energies, the pressure decreases sharply with distance from
the Galactic centre. For $E_0
> 1$~GeV, diffusion keeps the electron-positron spectrum (and the ensuing
pressure) roughly constant within a radius of a few kpc. The highest
values of the dark matter pressure are found for an initial energy $E_0 = 100\
\rm{MeV}$. Although the spectrum for $E_0 = 1$ and 10~MeV is
similar (approximately constant up to the cutoff at $\gamma_0$, because it
dominated by ionization losses) and even has a higher normalization
near the centre, the smaller upper limit of the integral in~(\ref{eqPressure})
yields significantly lower pressures. At high ($E_0 \geq 1$~GeV)
injection energies, the dark matter pressure is also lower, due to the smaller
number density of dark matter particles. Most of the dark matter pressure for
an initial energy $E_0$, except $E_0= 1\ \rm{TeV}$, are higher
than the pressure from the gas in the galaxy and for $E_0 = 10\ \rm {MeV}- 1\
\rm{GeV}$, the dark matter pressure is significantly higher than the
pressure from magnetic fields.

%__________________________________
 \subsection{Astrophysical parameters}

We will now discuss the effect of the various astrophysical parameters that
enter our calculation of the dark matter pressure, namely the
intensity of the magnetic field, the density and ionization fraction of the ISM
gas, and the inner slope of the dark matter density profile. As
we did for the canonical model, we will compare the results obtained for
different initial energies $E_0$ from $\sim 1$~MeV to 1~TeV and vary
each of the astrophysical parameters in turn in order to assess how much they
influence the results.

Magnetic fields affect the high-energy ($\gamma>10^3$) tail of the
electron-positron spectrum by setting the energy losses due to synchrotron
radiation. As can be seen in Figure~\ref{figLoss}, in our canonical model with
$B = 6~\mu$G, the synchrotron term~(\ref{eqsynloss}) is
responsible for about 50 per cent of the energy loss at high energies, with ICS
being responsible for most of the other 50 per cent. At low
energies, energy losses are dominated by ionization of neutral hydrogen, and the
contribution of synchrotron emission is negligible. The effect
of varying $B$ from 1 to 10~$\mu$G is plotted on the left panel in
Figure~\ref{figAstro}. Not surprisingly, the results for an initial energy
$E_0<1\ \rm{GeV}$ are largely unaffected. At higher energies, the pressure at
large radii decreases with the magnetic field intensity because of
the more rapid energy losses. However, the diffusion length becomes shorter,
increasing the effective production rate and yielding a larger
pressure near the centre of the Galaxy.

%__________________________________
\begin{figure*}
\includegraphics[width=8cm]{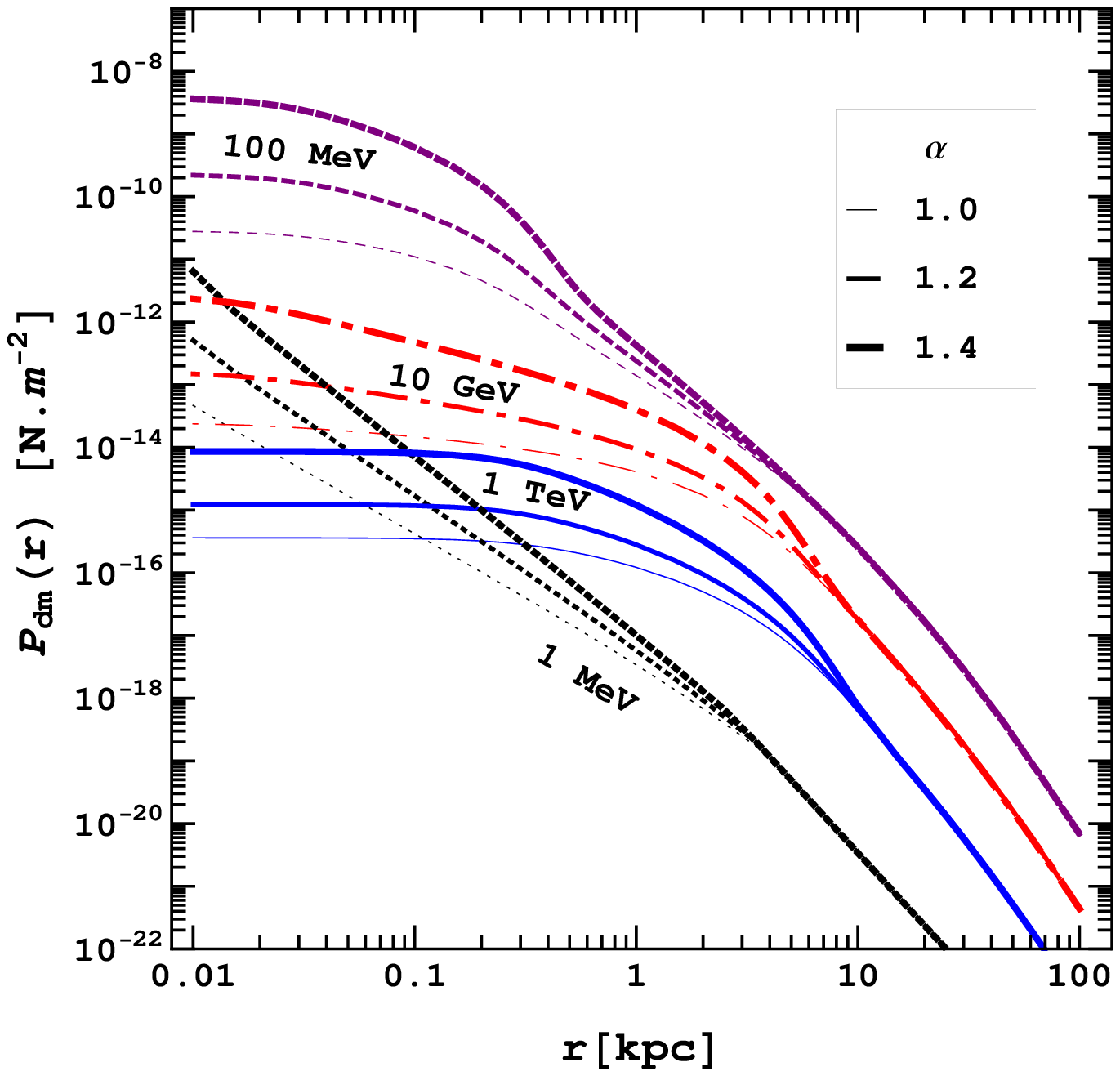}
~~\includegraphics[width=8cm]{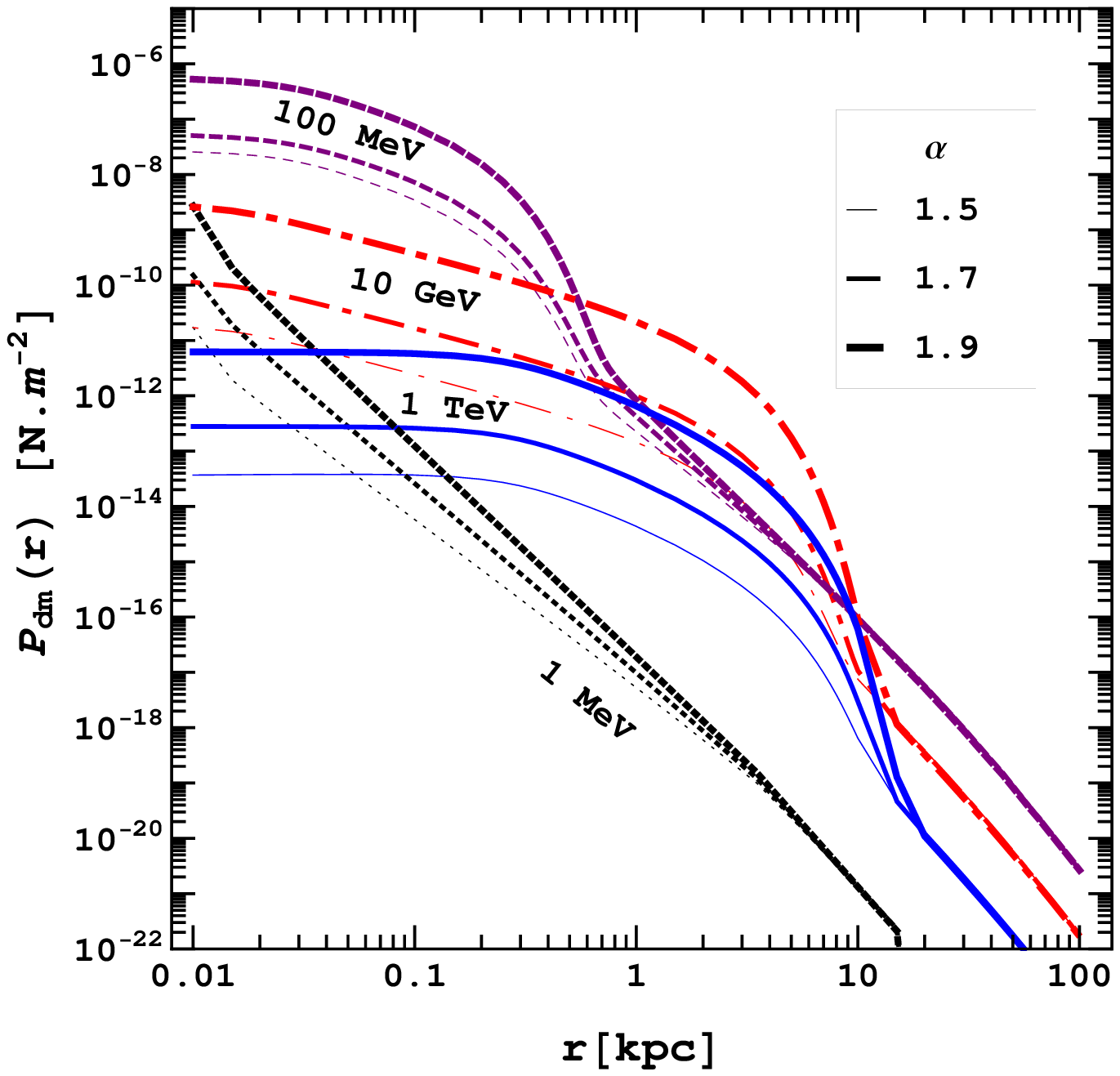} \caption{Dark matter
pressure for different values of $E_0$ and the inner logarithmic
slope $\alpha$ of the dark matter density profile.} \label{figAlpha}
\end{figure*}
%__________________________________

%__________________________________
\begin{table}
\begin{center}
\begin{tabular}{ccc}
$\alpha$ & $\rho_{\rm s} c^2$~[GeV~cm$^{-3}$] & $r_{\rm s}$~[kpc]\\ \hline
1.0 & 0.3483  & 16.68\\
1.2 & 0.1975  & 20.34\\
1.4 & 0.09946 & 25.88\\
1.5 & 0.06658 & 29.81\\
1.7 & 0.02469 & 42.46\\
1.9 & 0.00615 & 70.30
\end{tabular}
\end{center}
\caption{Characteristic density and radius of the dark matter density
profile~(\ref{eqDMProfile1}) as a function of its asymptotic logarithmic inner
slope $\alpha$.}
\label{tabRhosRs}
\end{table}
%__________________________________

%__________________________________
\begin{figure*}
\includegraphics[width=5.5cm]{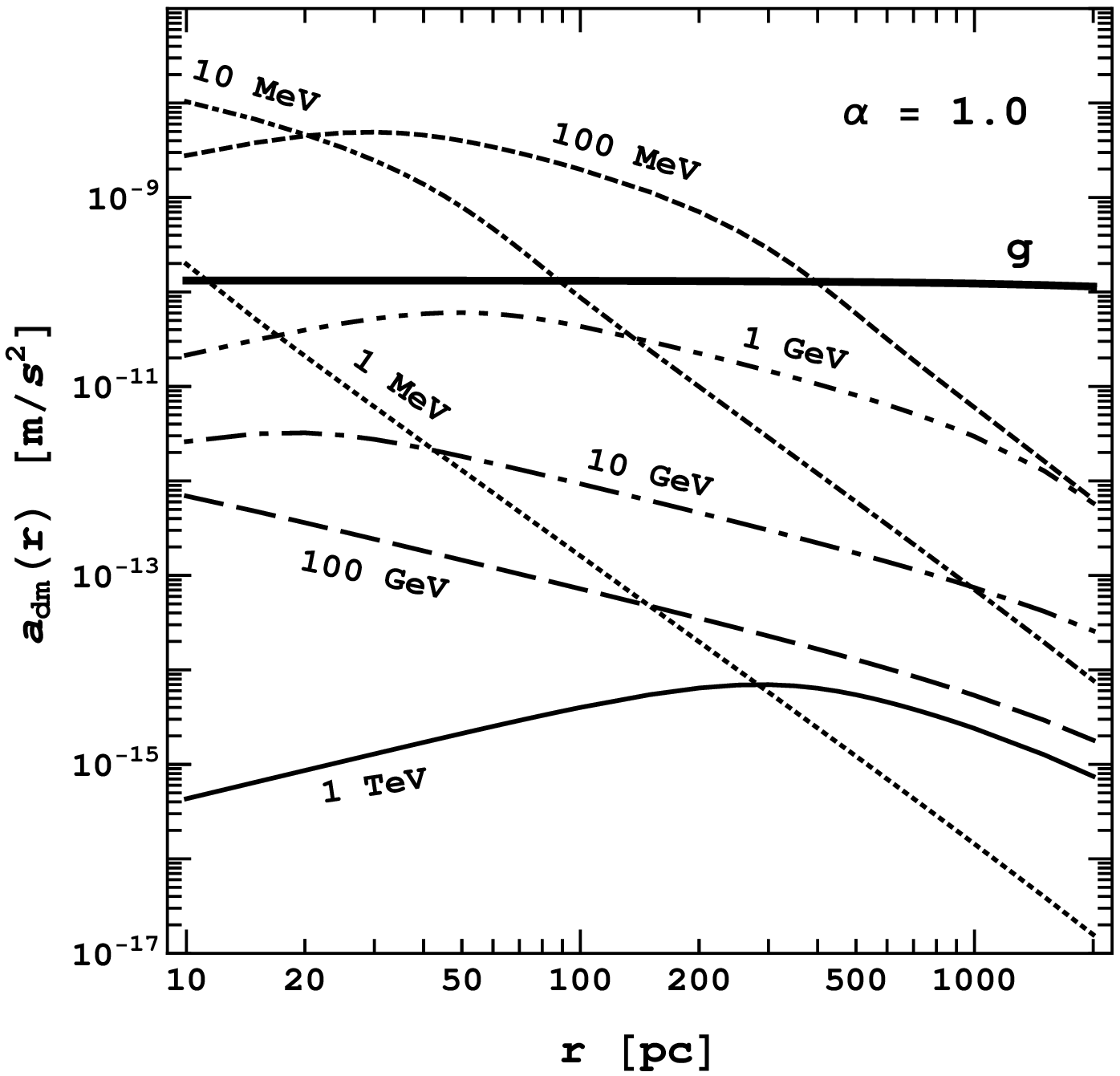}
\hfill \includegraphics[width=5.5cm]{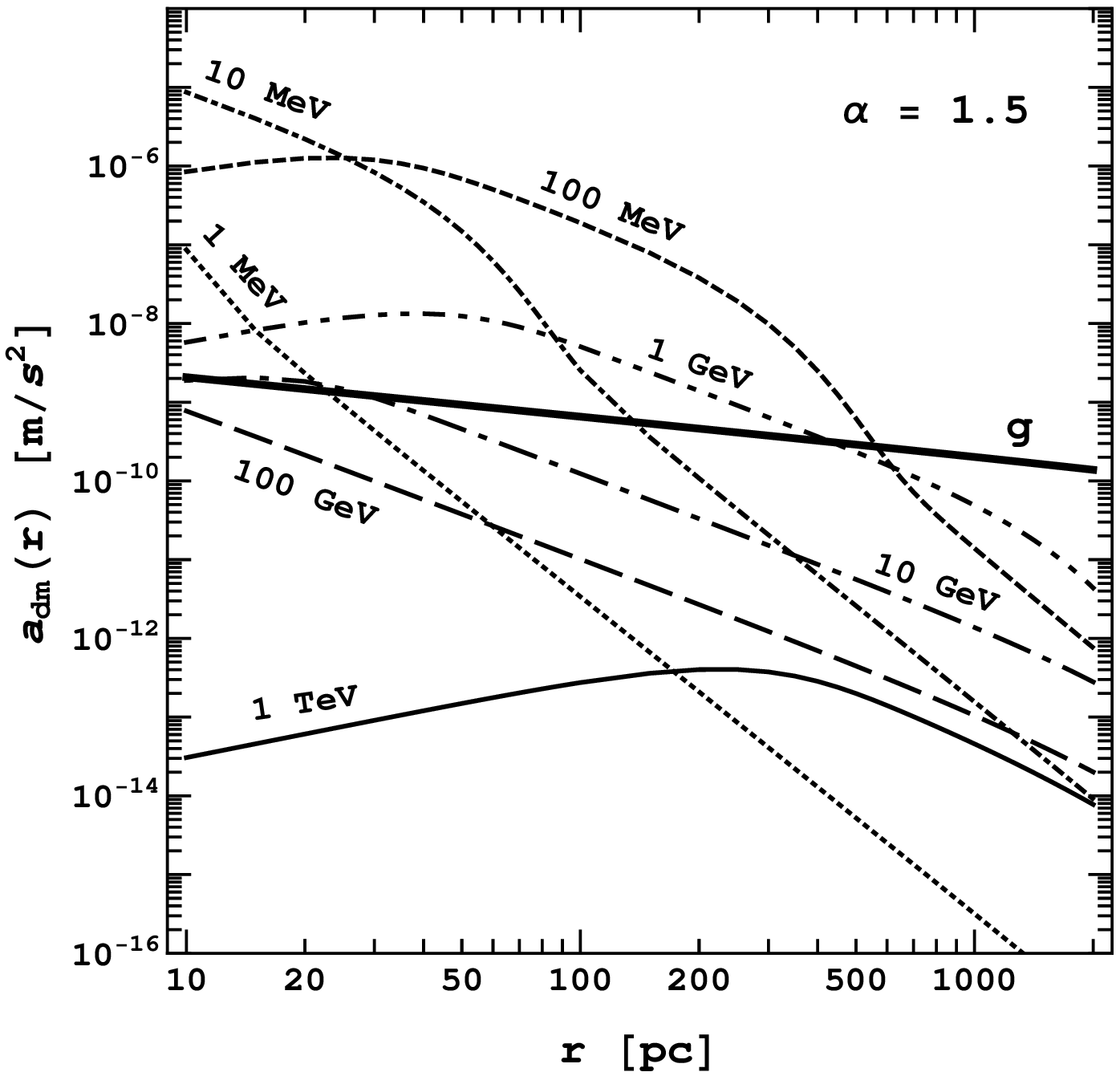}
\hfill \includegraphics[width=5.5cm]{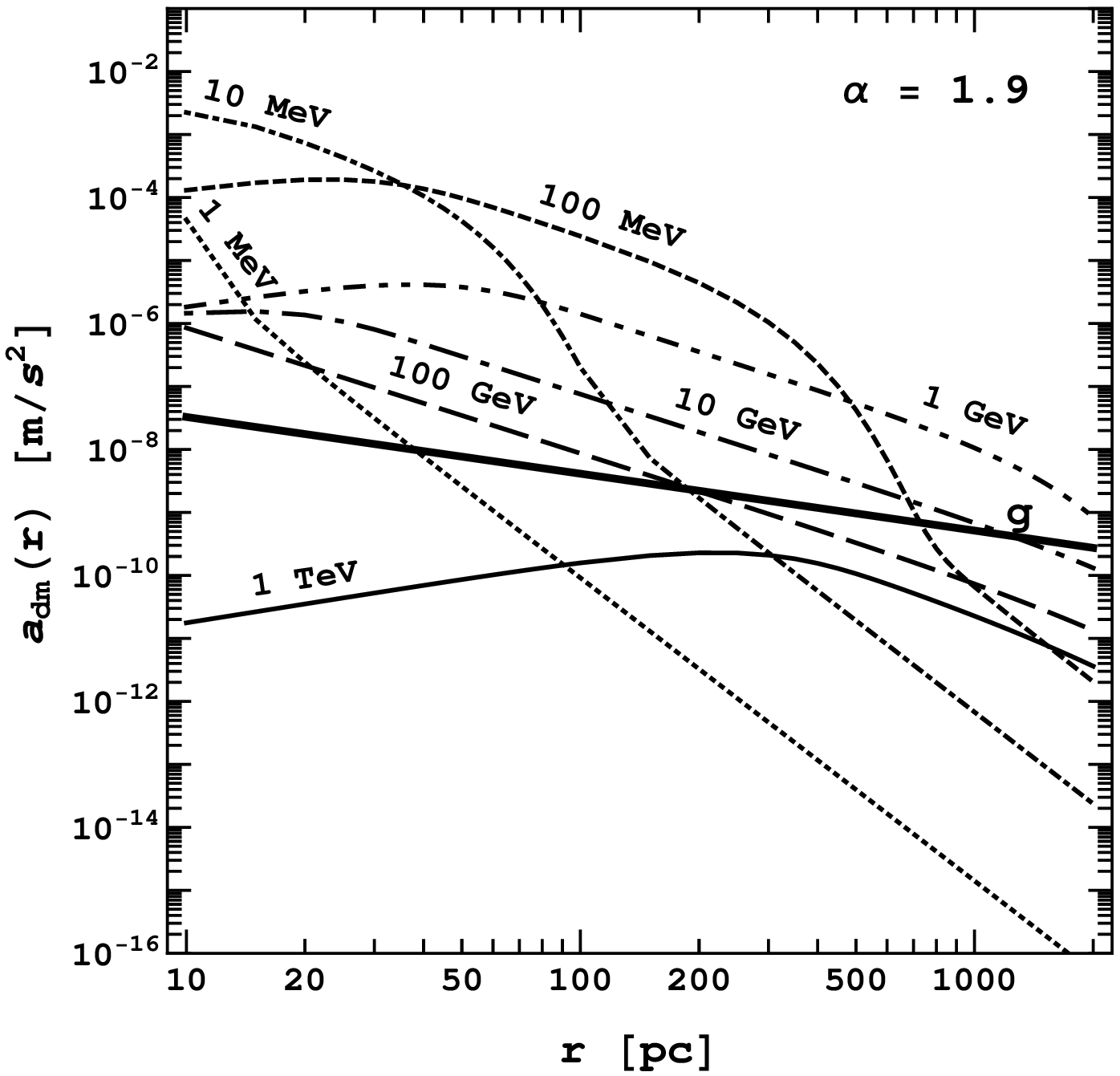}
\caption{Dark matter acceleration for different values of the injection energy
$E_0$ and the inner logarithmic slope $\alpha$.}
\label{figAcceleration}
\end{figure*}
%__________________________________

As explained in Section~\ref{secLoss}, the density of the ISM gas and the
ionization fraction $X_{\rm ion}$ regulate the energy losses by
Coulomb interactions, bremsstrahlung, and ionization. For $X_{\rm ion}=0$ (our
canonical model), the ISM gas is entirely composed of neutral
hydrogen atoms, and the energy loss of the electrons and positrons with
$\gamma<10^3$ is dominated by the ionization process. At the other
extreme, $X_{\rm ion}=1$, the ISM is already fully ionized, and the relevant
energy losses are Coulomb collisions and bremsstrahlung. Since the
total loss by these processes is higher than the loss by ionization, the maximum
pressure happens when $X_{\rm {ion}}=0$ (mid panel of
Figure~\ref{figAstro}). The effect of changing the gas density from $0.1$ to
$10\ \rm {cm^{-3}}$ is shown on the right panel. Higher densities
yield lower dark matter pressures, simply because the energy losses are faster.

Finally, we calculate the dependence of dark matter pressure on the inner
logarithmic slope $\alpha$ of the dark matter density profile. When
varying $\alpha$, we also modify the characteristic density and radius in
expression~(\ref{eqDMProfile1}) so that the dark matter density at the
solar radius is equal to 0.3~GeV~cm$^{-3}$ and the virial mass of the Galaxy is
$10^{12}$~M$_\odot$. The appropriate values of $\rho_{\rm s}$
and $r_{\rm s}$ are quoted in Table~\ref{tabRhosRs} for several values of the
inner slope. For $\alpha\geq 1.5$, the production rate $Q_0$ in
equation~(\ref{eqQ0}) diverges rapidly at $r=0$, so we add a cutoff based on the
local annihilation rate $\Gamma(r)$,
\be
Q_0(r) = 2\, \left[
\frac{ \rho_{\rm dm}(r) \exp(-t_0\Gamma(r)) }{ m_{\rm dm} } \right]^2 \langle
\sigma v \rangle_{e^\pm} \label{eqQ02}
\ee
where $t_0 = 13.7$~Gyr
is the age of the universe and $\Gamma(r) = \frac{ \rho_{\rm dm}(r) }{ m_{\rm
dm} } \langle \sigma v \rangle_{e^\pm}$.

The dark matter pressure profiles obtained for
$\alpha=1$, 1.2, 1.4, 1.5, 1.7 and 1.9 are compared in Figure~\ref{figAlpha}.
Since the central dark matter density increases dramatically with the value of
the inner logarithmic slope, this is, by far, the most relevant astrophysical
parameter, only second in importance to the injection energy $E_0$ related to
the mass (and the precise nature) of the dark matter particle.

%__________________________________
\begin{figure*}
\includegraphics[width=5.5cm]{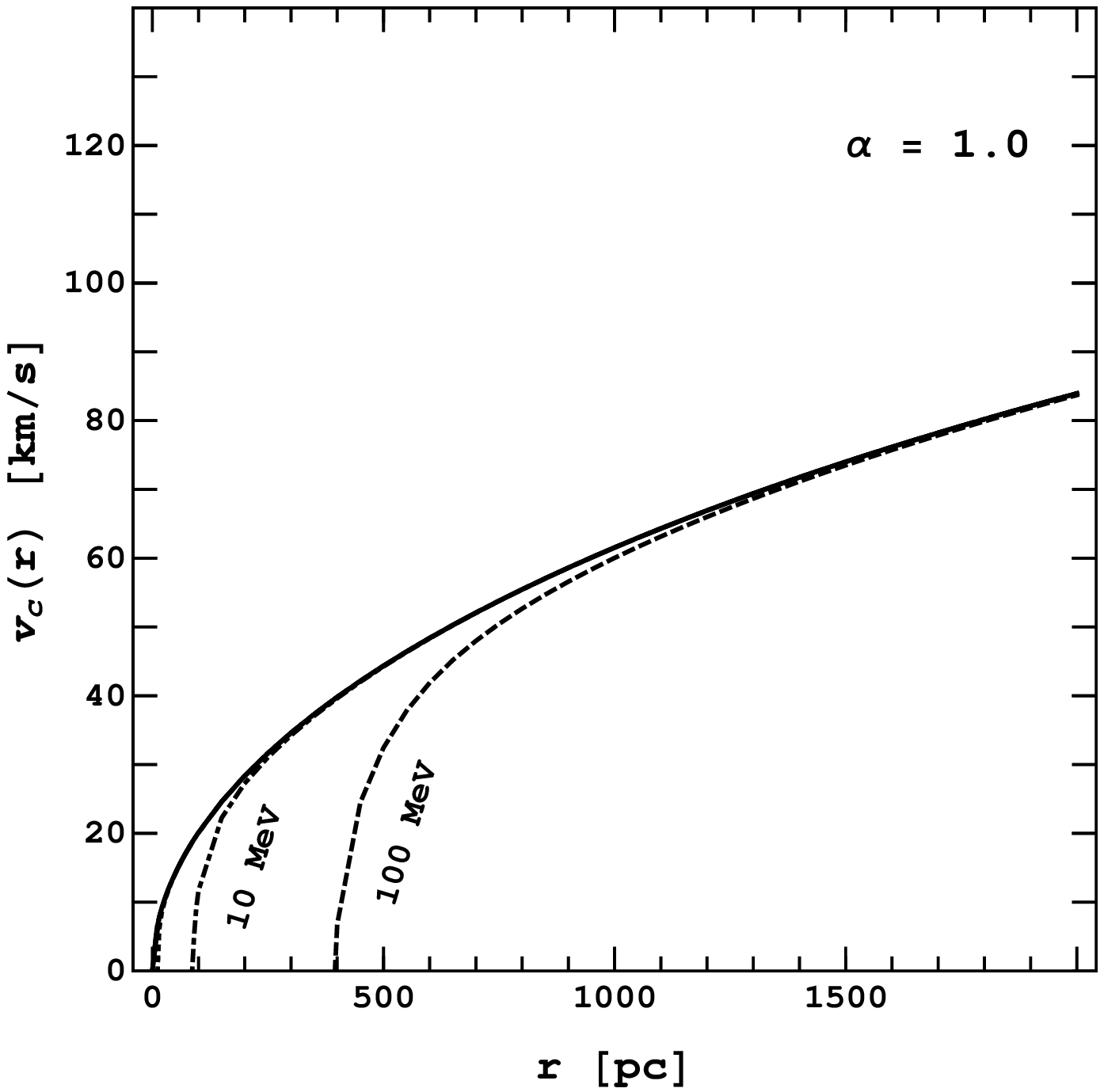}
\hfill \includegraphics[width=5.5cm]{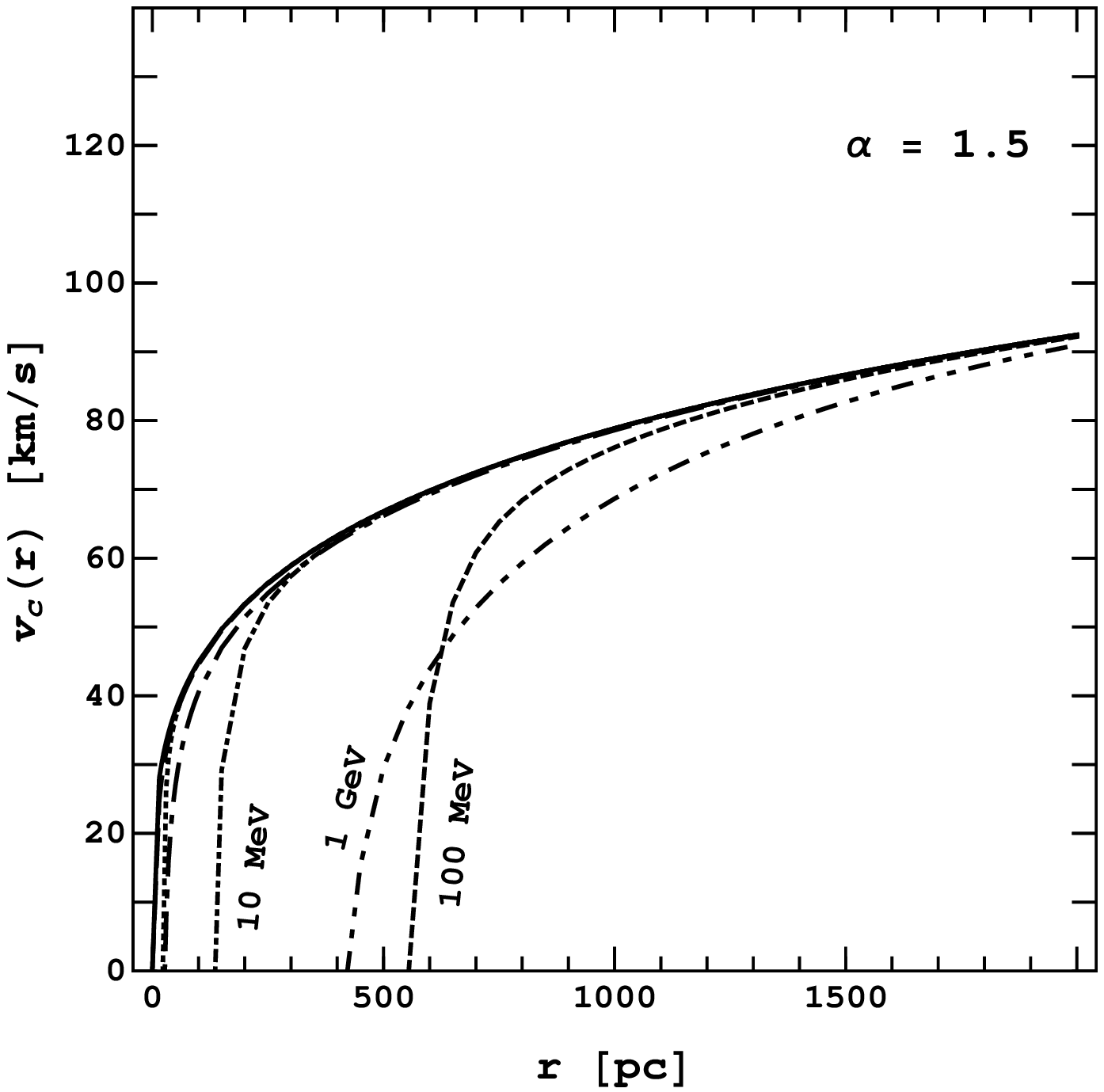}
\hfill \includegraphics[width=5.5cm]{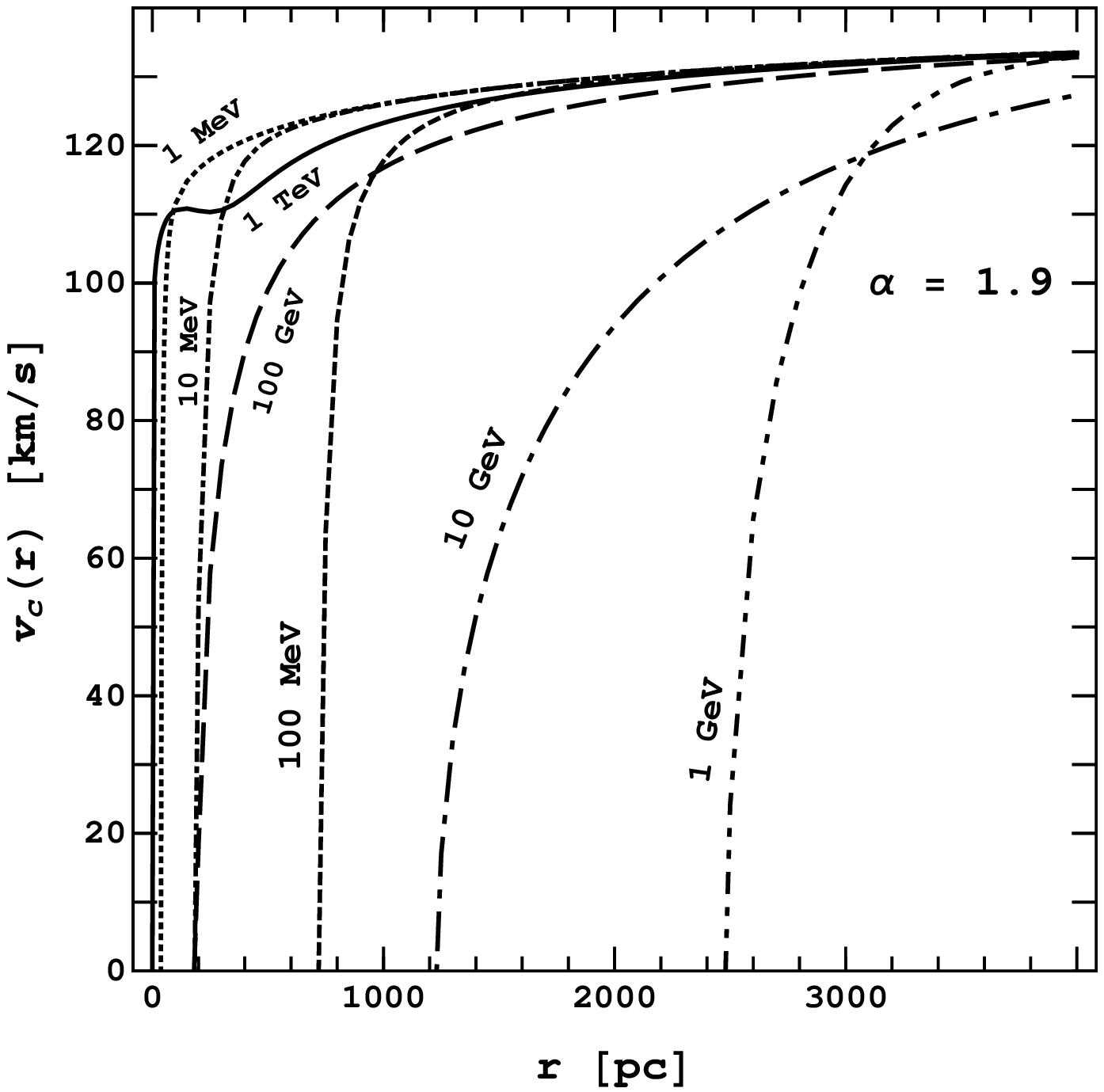}
\caption{Effect of dark matter annihilation on the rotation curve. From left to
right, panels correspond to $\alpha=1.0$ (our canonical model),
1.5, and 1.9, respectively.} \label{figVelocity}
\end{figure*}
%__________________________________

%--------------------------------------------------------------------------
 \section{Rotation curves}
 \label{ModelRotationCurves}
%--------------------------------------------------------------------------

The gradient of the dark matter pressure induces an acceleration on the baryonic
component that opposes the gravitational force. This
acceleration, given by expression~(\ref{eqAcceleration}), is plotted in
Figure~\ref{figAcceleration} for all the injection energies considered
in this work and compared to the gravitational acceleration $g(r) = GM(r)/r^2$
(represented by a thick solid black line). Each panel corresponds
to a different value of the inner logarithmic slope of the dark matter density
profile. For our canonical model with $\alpha=1$, $g(r) \approx
2\pi G \rho_{\rm s} r_{\rm s} = 1.3 \times 10^{-10}$~m~s$^{-1}$ in the innermost
regions, whereas in the general case described by
equation~(\ref{eqDMProfile1}) gravity scales as $g(r) \propto r^{1-\alpha}$ for
$r\ll r_{\rm s}$.

%__________________________________
\begin{table*}
\begin{center}
\begin{tabular}{cccccccc}
Name & $\rho_{\rm s} c^2$~[GeV~cm$^{-3}$] & $r_{\rm s}$~[kpc] &
$\chi^2_{\rm{dBB}}$ &  $\chi^2_{\rm{100MeV, HI}}$ &$\chi^2_{\rm{1GeV,
HI}}$ & $\chi^2_{\rm{100MeV, HII}}$ & $\chi^2_{\rm{1GeV, HII}}$
\\\hline
UGC\,1230 & $9.2799 \times 10^{-2}$ & 15.152 & 0.723 & 0.713 & 0.720 &  49.825 &
0.430  \\
UGC\,5005 & $8.5056 \times 10^{-5}$ & 4206.9 & 0.521 & 0.520 & 0.521 &   4.726 &
0.488  \\
LSBC\,F563-01&$4.5080\times10^{-3}$ & 188.61 & 0.383 & 0.381 & 0.383 &  27.368 &
0.345  \\
UGC\,4173 & $8.5056 \times 10^{-5}$ & 1765.8 & 0.225 & 0.225 & 0.225 &   1.223 &
0.214  \\
UGC\,3371 & $8.5056 \times 10^{-5}$ & 7302.7 & 0.387 & 0.383 & 0.386 &  17.775 &
0.273  \\
NGC\,1560 & $8.5056 \times 10^{-5}$ & 7131.5 & 8.914 & 9.183 & 8.896 & 659.459 &
8.720  \\
DDO\,189  & $3.9944 \times 10^{-3}$ & 135.71 & 0.173 & 0.171 & 0.173 &  40.124 &
0.135  \\
NGC\,4395 & $3.8588 \times 10^{-1}$ & 5.8478 & 0.573 & 1.627 & 0.555 & 743.952 &
14.954 \\
NGC\,3274 & $4.4477 \times 10^{-1}$ & 6.7189 & 1.787 & 7.139 & 1.731 &2935.570 &
33.516 \\
NGC\,4455 & $8.5056 \times 10^{-5}$ & 8252.1 & 1.428 & 1.694 & 1.419 & 320.260 &
1.643  \\
NGC\,2366 & $1.7431 \times 10^{-1}$ & 2.3734 & 1.246 & 1.209 & 1.245 &  58.692 &
1.204  \\
UGC\,4325 & $8.5056 \times 10^{-5}$ & 23618  & 1.326 & 1.294 & 1.300 & 748.686 &
4.459  \\
DDO\,47   & $8.5056 \times 10^{-5}$ & 14107  & 0.400 & 0.303 & 0.394 & 260.841 &
2.056  \\
DDO\,185  & $8.5056 \times 10^{-5}$ & 5245.2 & 2.036 & 2.024 & 2.026 & 141.203 &
1.896  \\
\end{tabular}
\end{center}
\caption
{
\Referee
{
Galaxy name, characteristic dark matter density and radius, and reduced $\chi^2$
values for the maximum disk models in \citet{deBlokBosma02}.
$\chi^2_{\rm{dBB}}$ refers to the original model without dark matter
annihilation.
The subscripts in the other columns denote the injection energy (100~MeV or
1~GeV) and the conditions in the baryonic medium ($n=1$~cm$^{-3}$ and $x_{\rm
ion} = 0$ for HI, $n = 0.01$~cm$^{-3}$ and $x_{\rm ion} = 1$ for HII).
}
}
\label{tabGalaxies_maxDisk}
\end{table*}
%__________________________________

%__________________________________
\begin{table*}
\begin{center}
\begin{tabular}{cccccccc}
Name & $\rho_{\rm s} c^2$~[GeV~cm$^{-3}$] & $r_{\rm s}$~[kpc] &
$\chi^2_{\rm{dBB}}$ &  $\chi^2_{\rm{100MeV, HI}}$ &$\chi^2_{\rm{1GeV,
HI}}$ & $\chi^2_{\rm{100MeV, HII}}$ & $\chi^2_{\rm{1GeV, HII}}$
\\\hline
UGC\,1230 & $3.7001 \times 10^{-1}$ & 9.0973 & 1.091 & 1.032 & 1.072 & 257.449 
& 5.447   \\
UGC\,5005 & $9.1650 \times 10^{-3}$ & 75.759 & 0.175 & 0.172 & 0.174 & 18.571  
& 0.102   \\
LSBC\,F563-01&$9.9553\times10^{-2}$ & 20.850 & 0.370 & 0.355 & 0.367 & 155.857 
& 1.492   \\
UGC\,4173 & $8.5056 \times 10^{-5}$ & 2413.7 & 0.124 & 0.123 & 0.124 & 3.687   
& 0.110   \\
UGC\,3371 & $8.5056 \times 10^{-5}$ & 10073  & 0.280 & 0.273 & 0.278 & 36.263  
& 0.118   \\
NGC\,1560 & $5.6447 \times 10^{-3}$ & 166.59 & 2.330 & 3.810 & 2.315 & 1905.660
& 17.378  \\
DDO\,189  & $2.2759 \times 10^{-3}$ & 7.4629 & 0.109 & 0.088 & 0.105 & 273.964 
& 5.120   \\
NGC\,4395 & $4.4477 \times 10^{-1}$ & 6.3057 & 0.644 & 2.476 & 0.613 & 1149.580
& 30.584  \\
NGC\,3274 & $2.2174 \times 10^{-0}$ & 2.9131 & 0.941 & 29.406& 0.808 & 9205.720
& 371.555 \\
NGC\,4455 & $8.5056 \times 10^{-5}$ & 9702.7 & 0.614 & 1.078 & 0.605 & 466.972 
& 2.482   \\
NGC\,2366 & $3.8588 \times 10^{-1}$ & 4.1902 & 1.935 & 2.222 & 1.900 & 950.701 
& 16.488  \\
UGC\,4325 & $8.5056 \times 10^{-5}$ & 36371  & 1.096 & 1.081 & 1.047 & 1737.360
& 34.844  \\
DDO\,47   & $8.5056 \times 10^{-5}$ & 16137  & 0.272 & 0.180 & 0.265 & 344.709 
& 3.186   \\
DDO\,185  & $8.5056 \times 10^{-5}$ & 7876.7 & 2.162 & 2.333 & 2.138 & 366.502 
& 3.996   \\
\end{tabular}
\end{center}
\caption
{
\Referee
{
Same as Table~\ref{tabGalaxies_maxDisk}, for the constant mass-to-light ratio
models in \citet{deBlokBosma02}.
}
}
\label{tabGalaxies_ML}
\end{table*}
%__________________________________

%__________________________________
\begin{figure*}
\includegraphics[width=8.8cm]{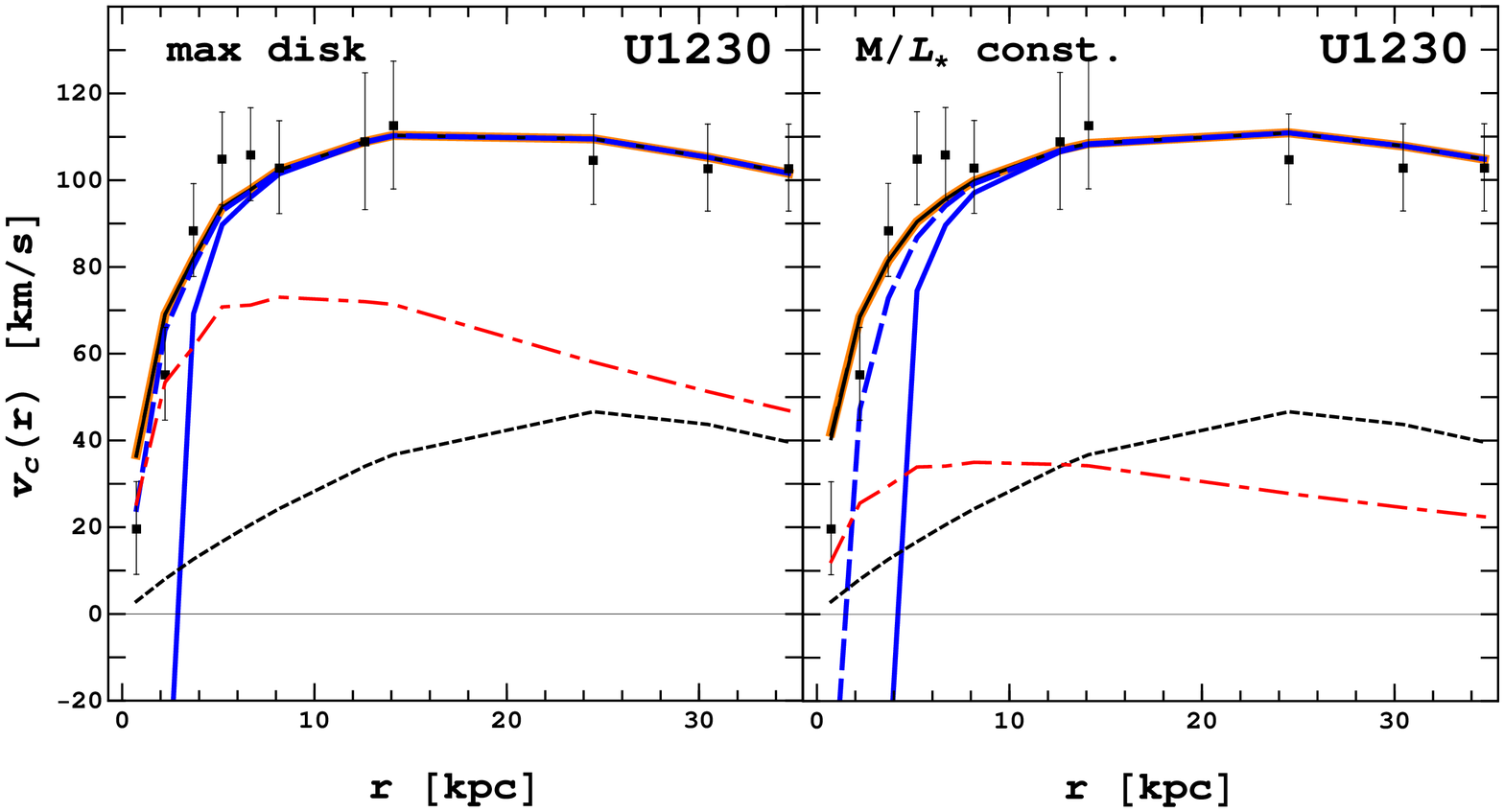} \hfill
\includegraphics[width=8.8cm]{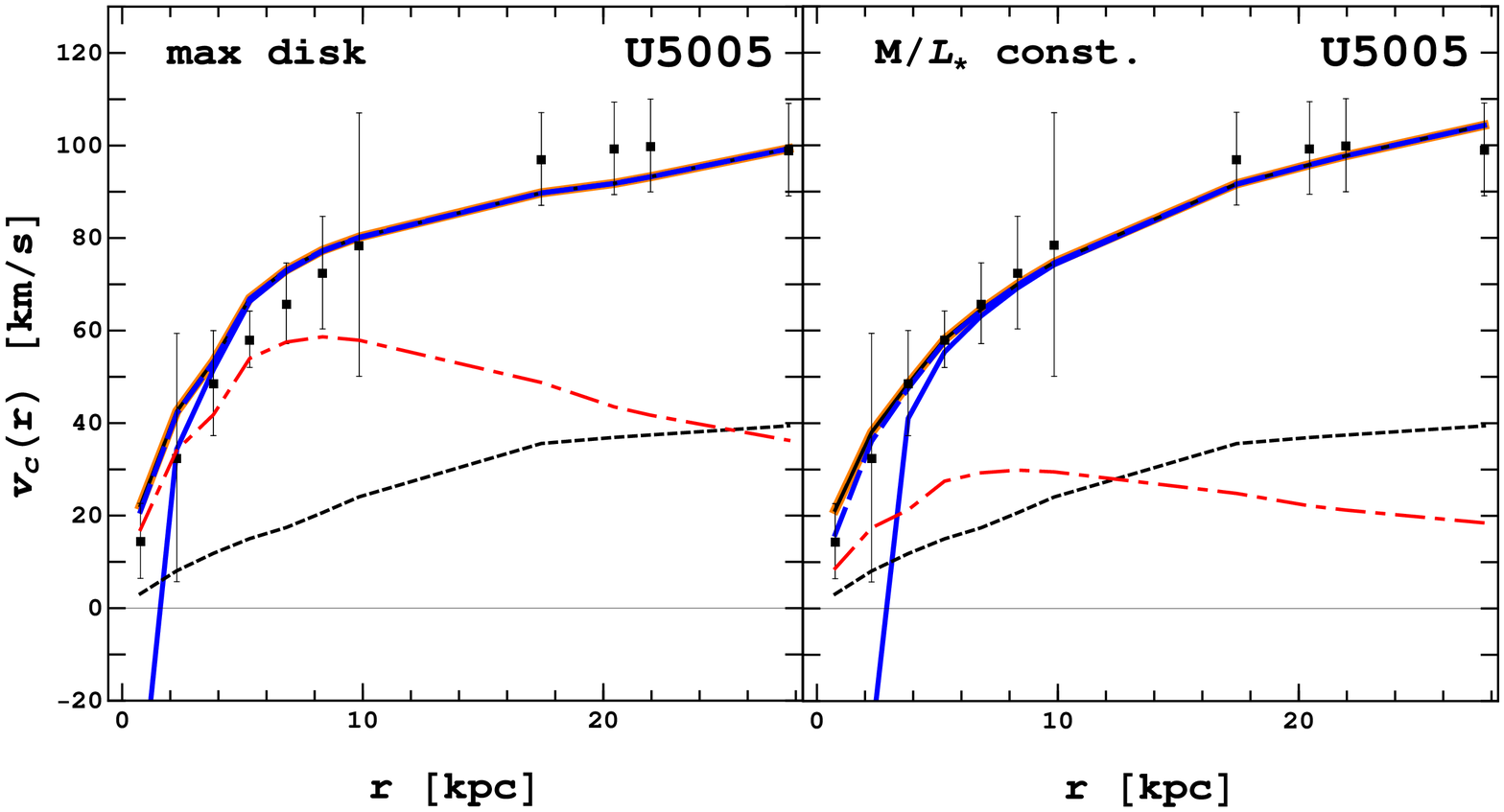} \hfill
\includegraphics[width=8.8cm]{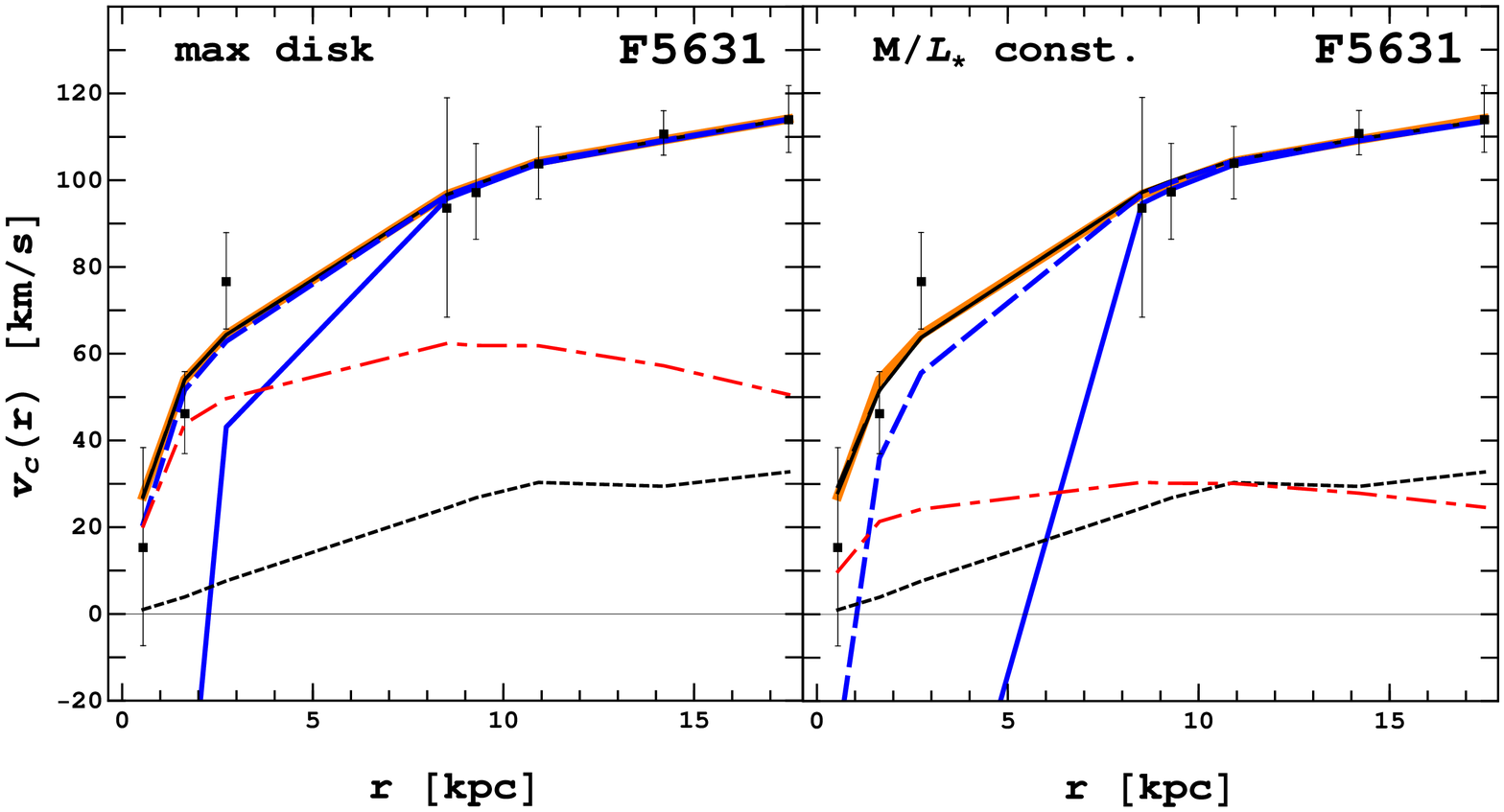} \hfill
\includegraphics[width=8.8cm]{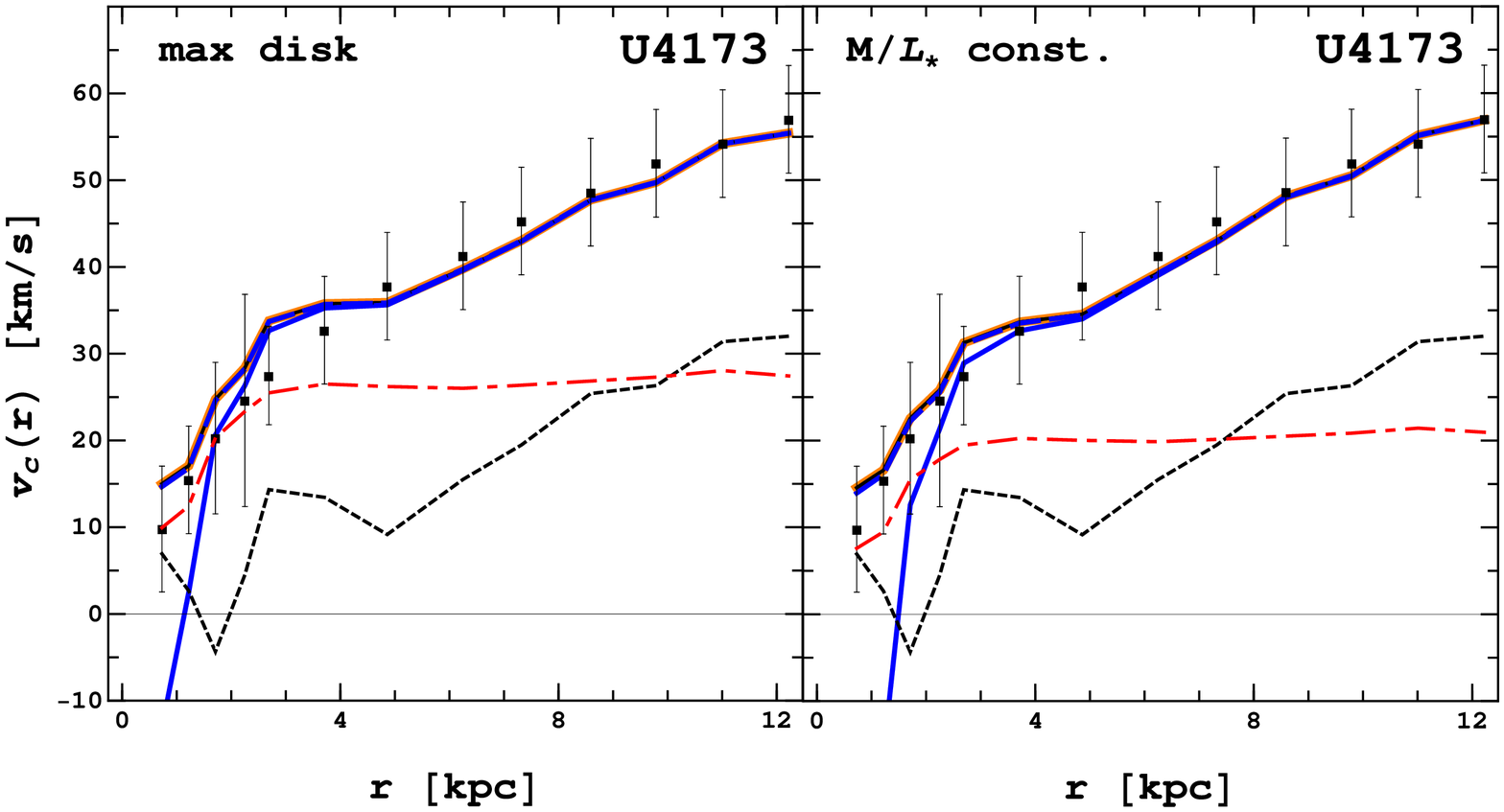} \hfill
\includegraphics[width=8.8cm]{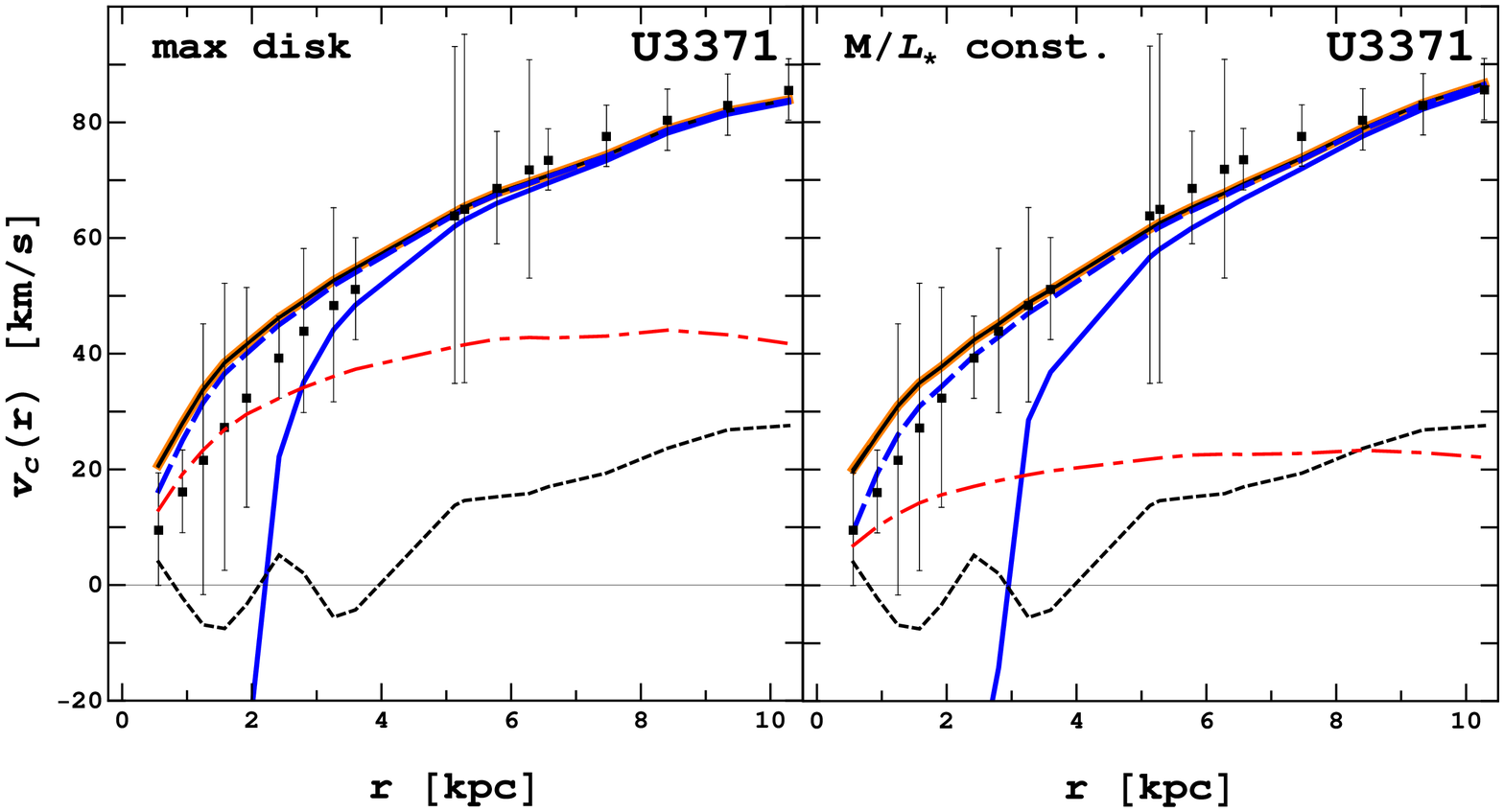} \hfill
\includegraphics[width=8.8cm]{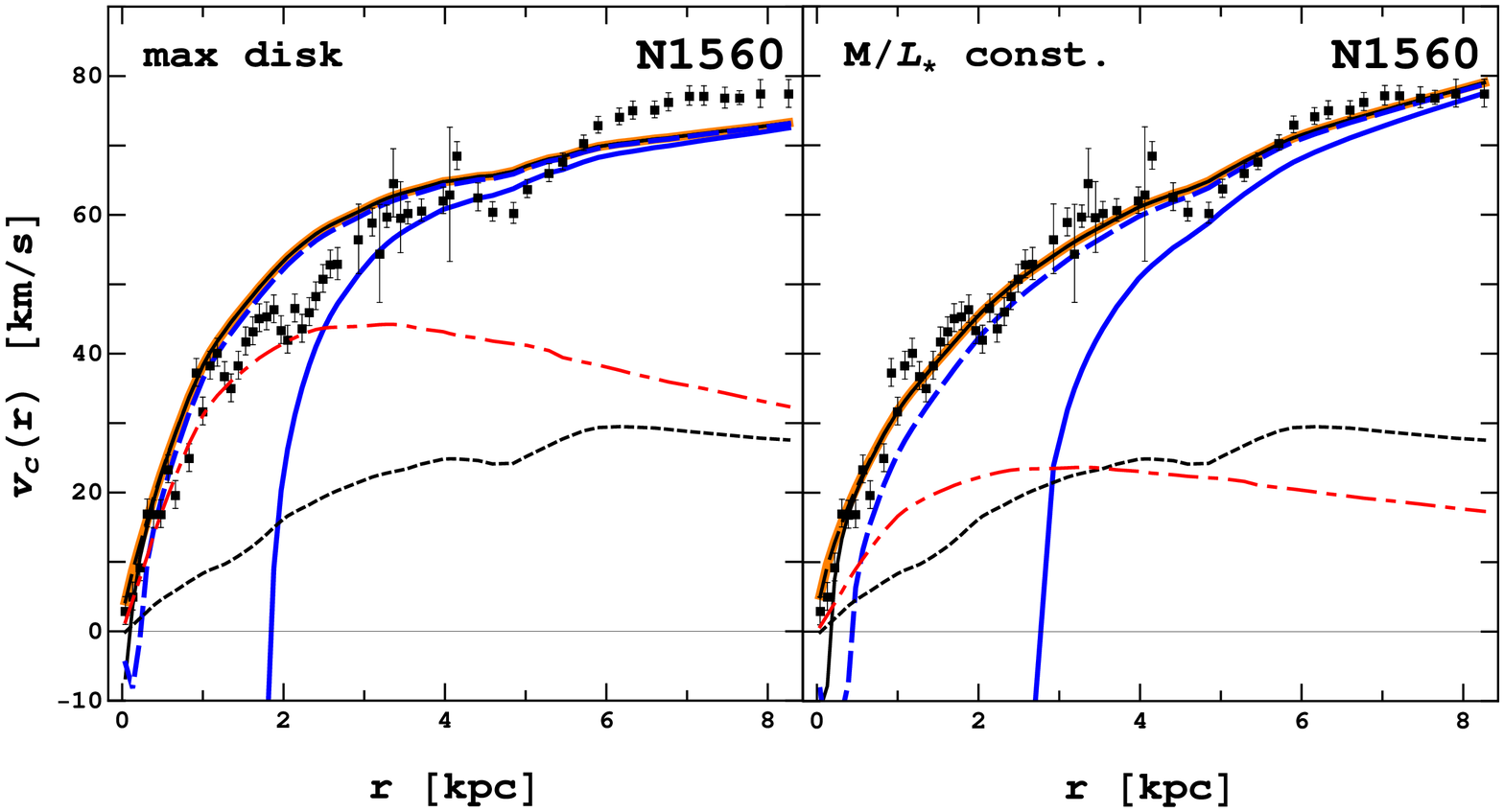} \hfill
\includegraphics[width=8.8cm]{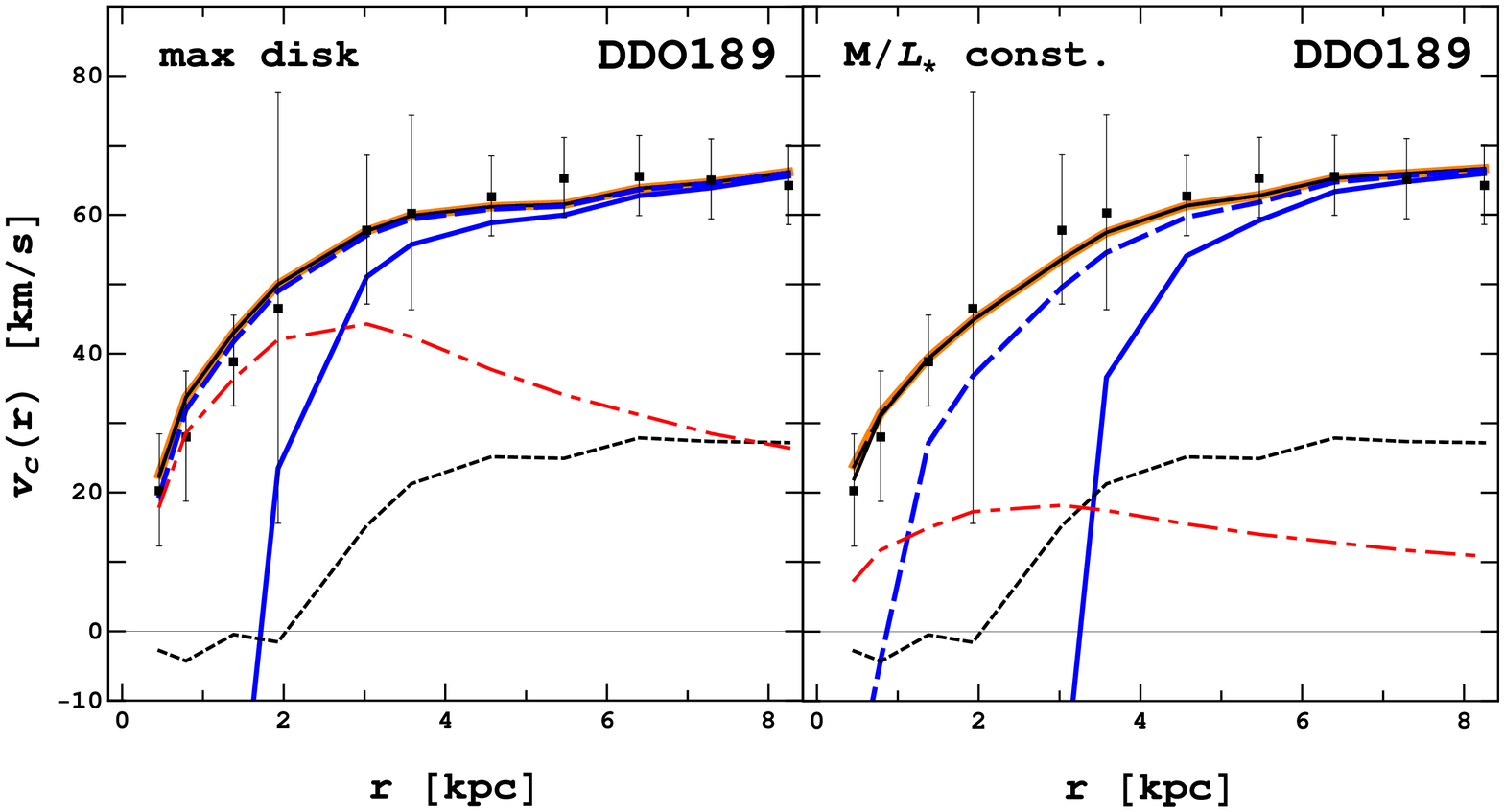}\hfill
\includegraphics[width=8.8cm]{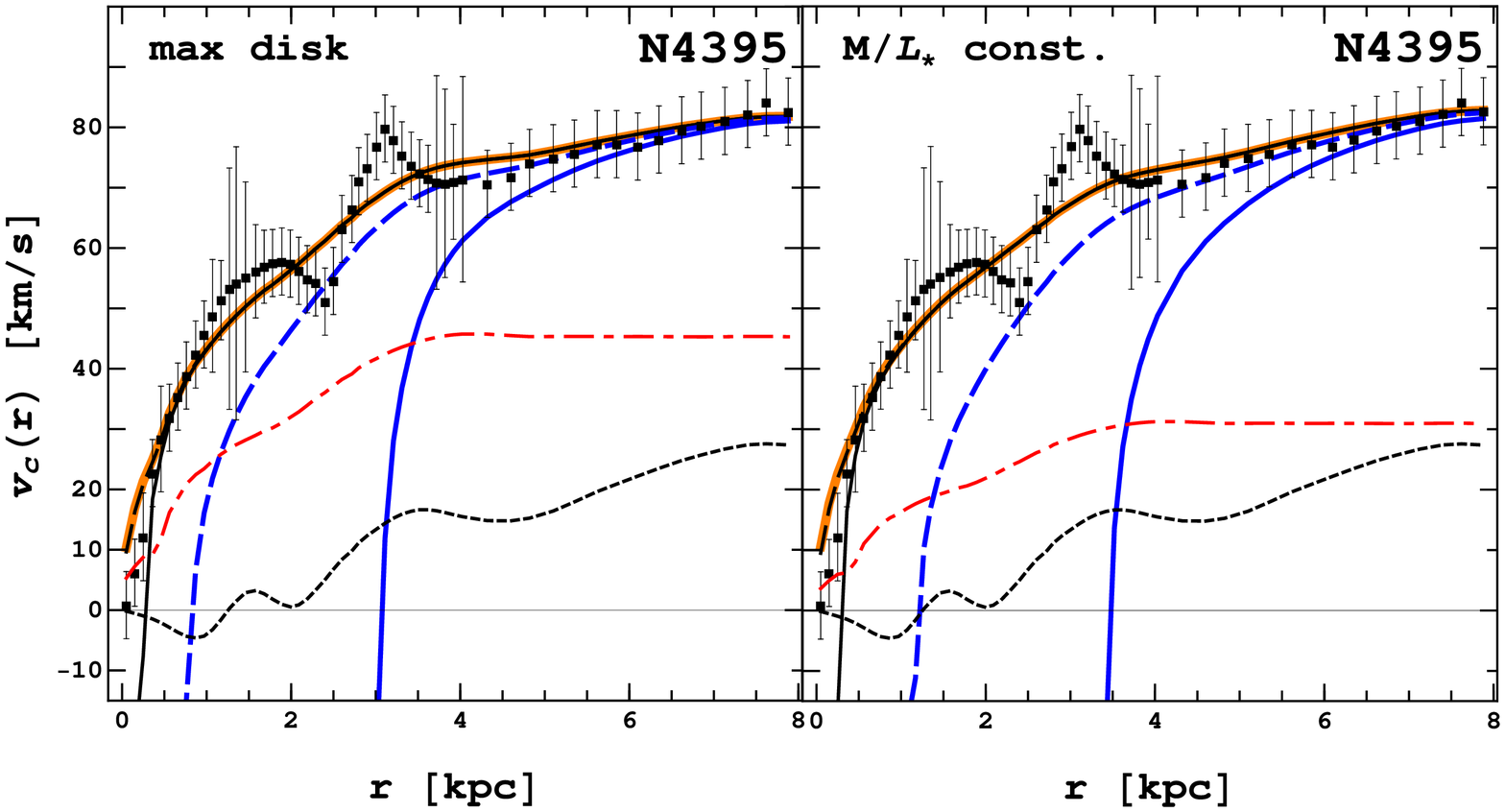} \hfill
\caption
{
\Referee
{
Model predictions, compared to the observed rotation curves (data points with
error bars).
Solid lines are used for $E_0=100$~MeV and dashed lines for 1~GeV.
Results for neutral and ionized gas are shown in black and blue colour,
respectively.
Orange lines depict the model without dark matter annihilation. The rotation
curves by star and gas \citep[data taken from][]{deBlokBosma02} are shown by red dot-dashed
and black small dashed lines, respectively.
}
}
\label{figGalaxies1}
\end{figure*}
%__________________________________

%__________________________________
\begin{figure*}
\includegraphics[width=8.8cm]{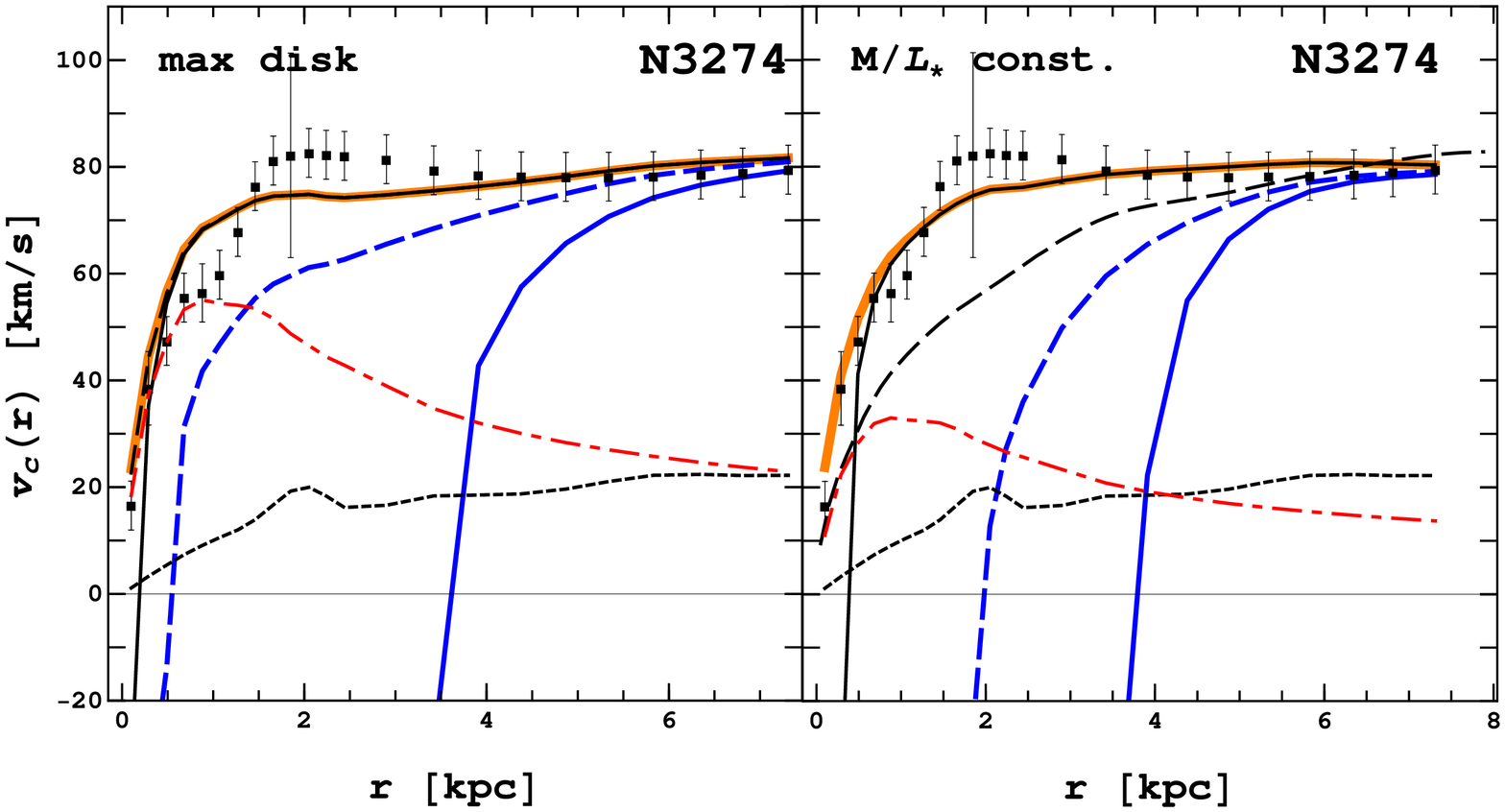} \hfill
\includegraphics[width=8.8cm]{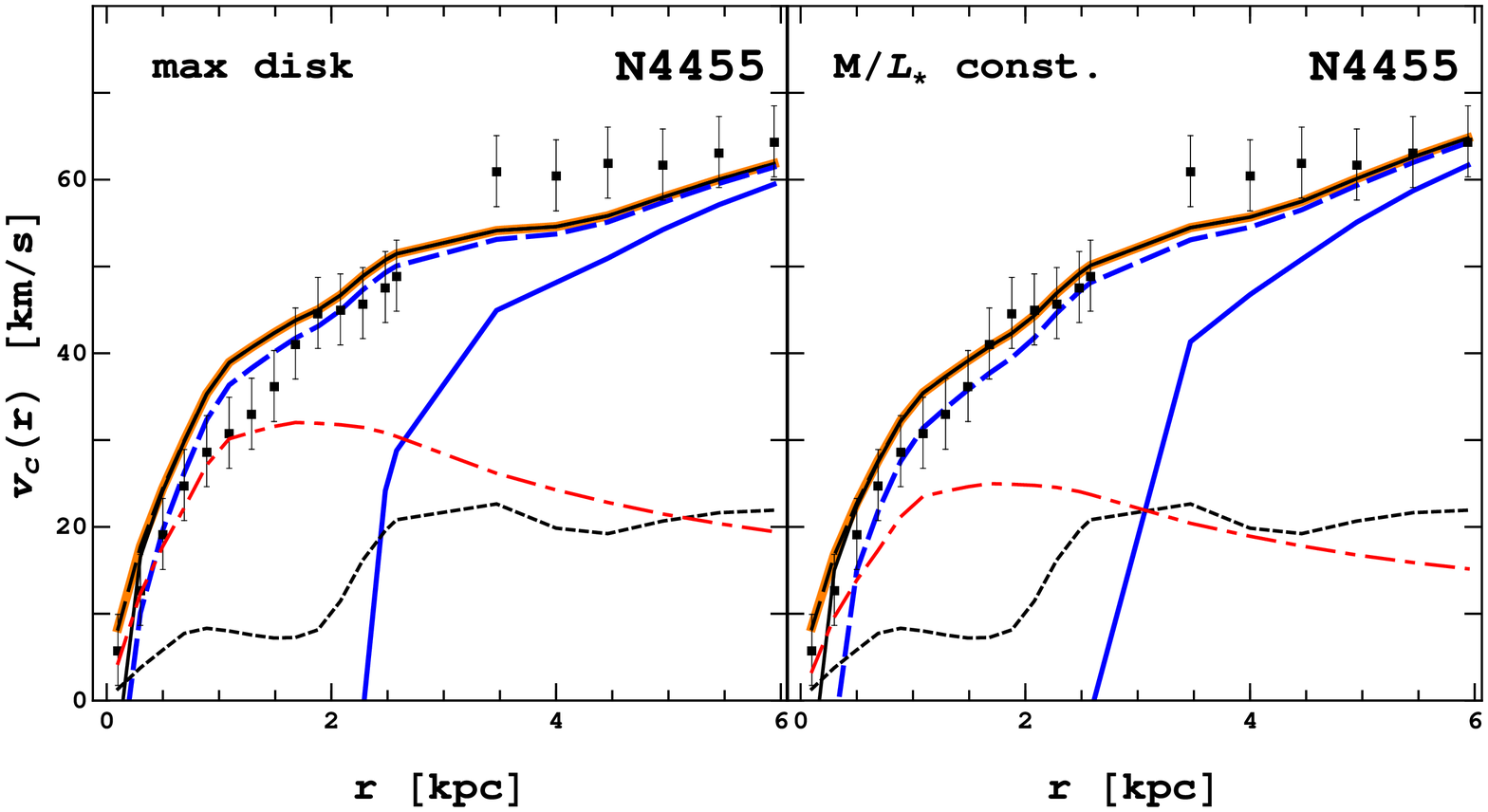} \hfill
\includegraphics[width=8.8cm]{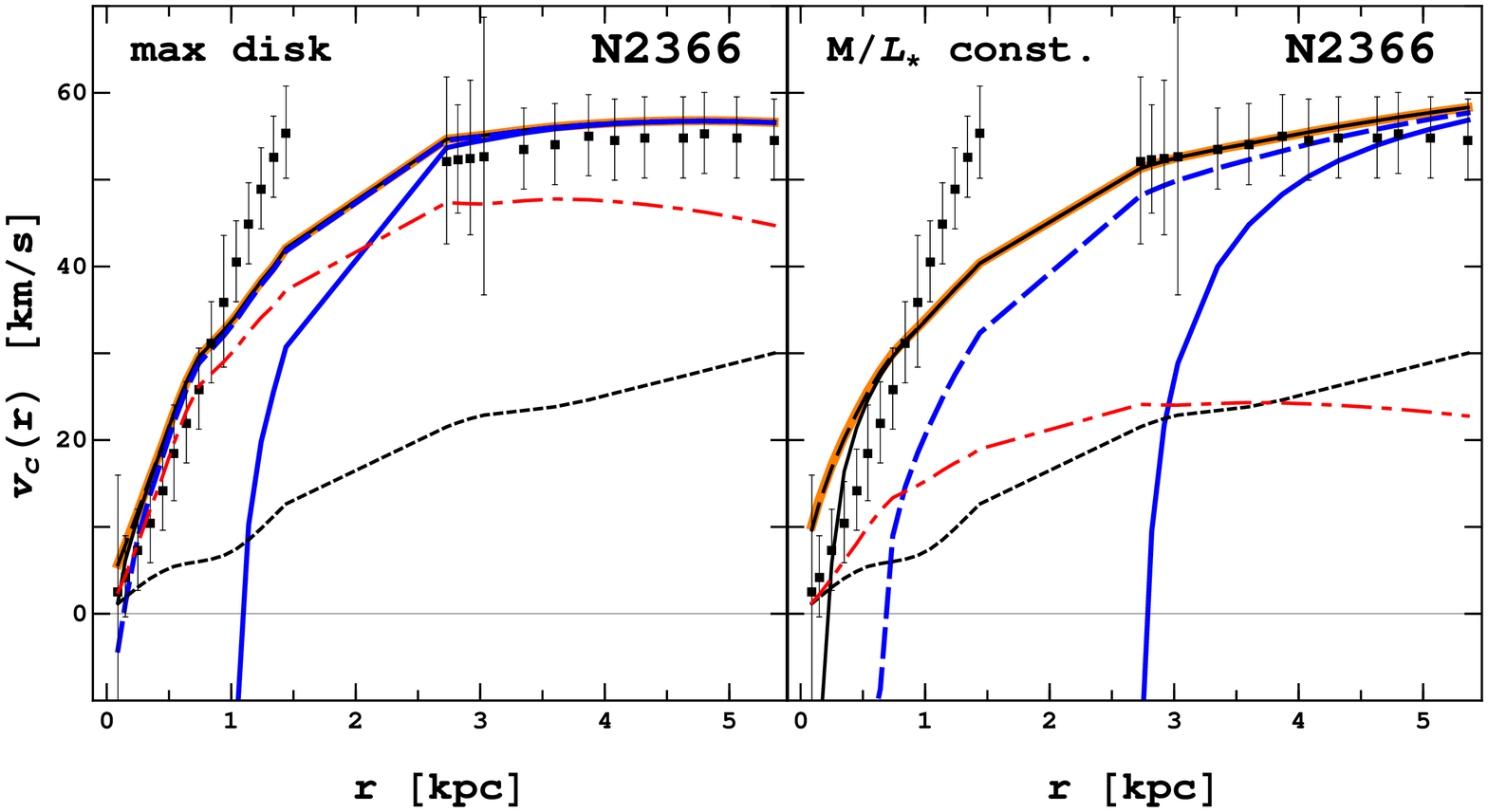} \hfill
\includegraphics[width=8.8cm]{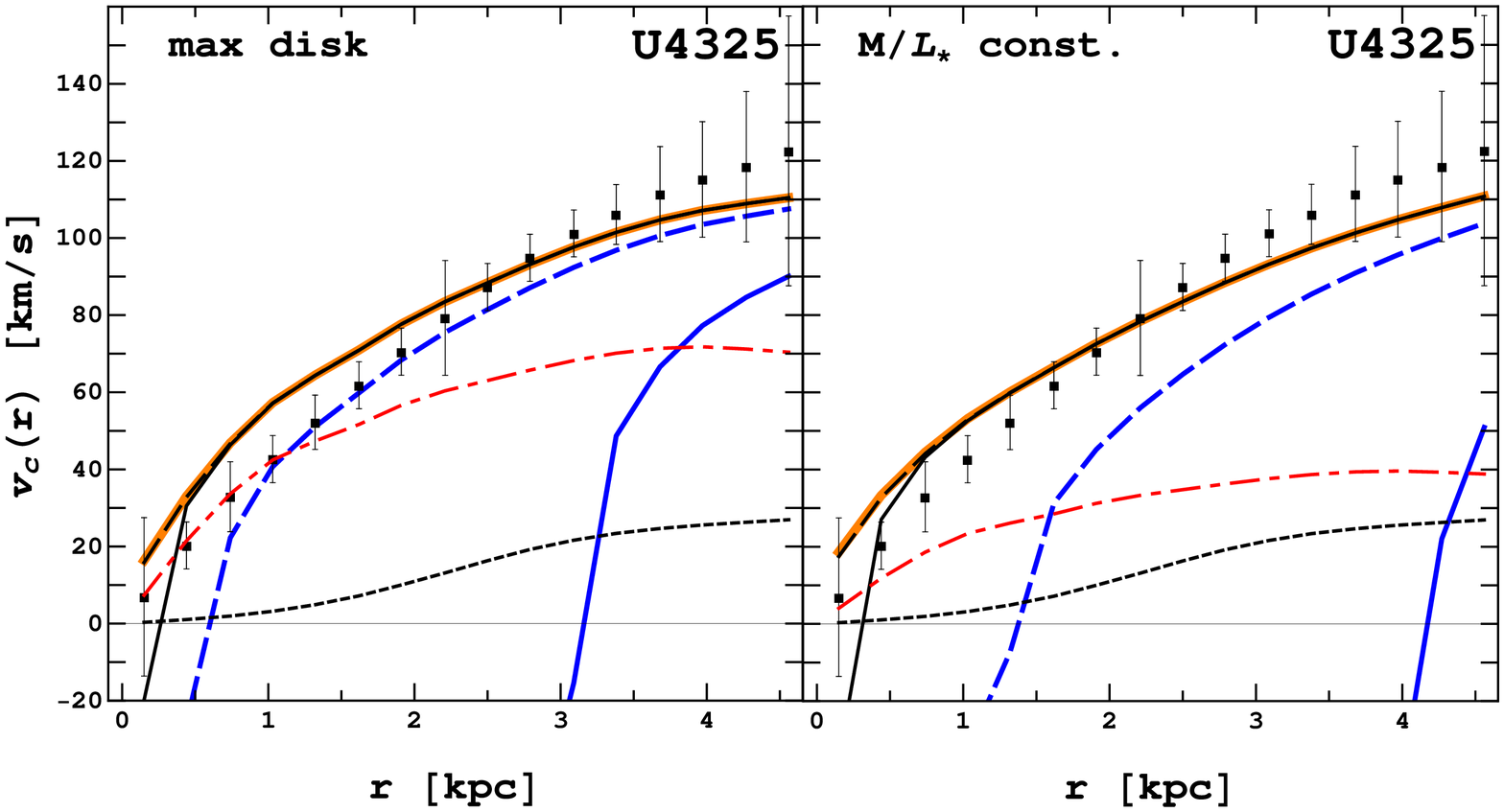} \hfill
\includegraphics[width=8.8cm]{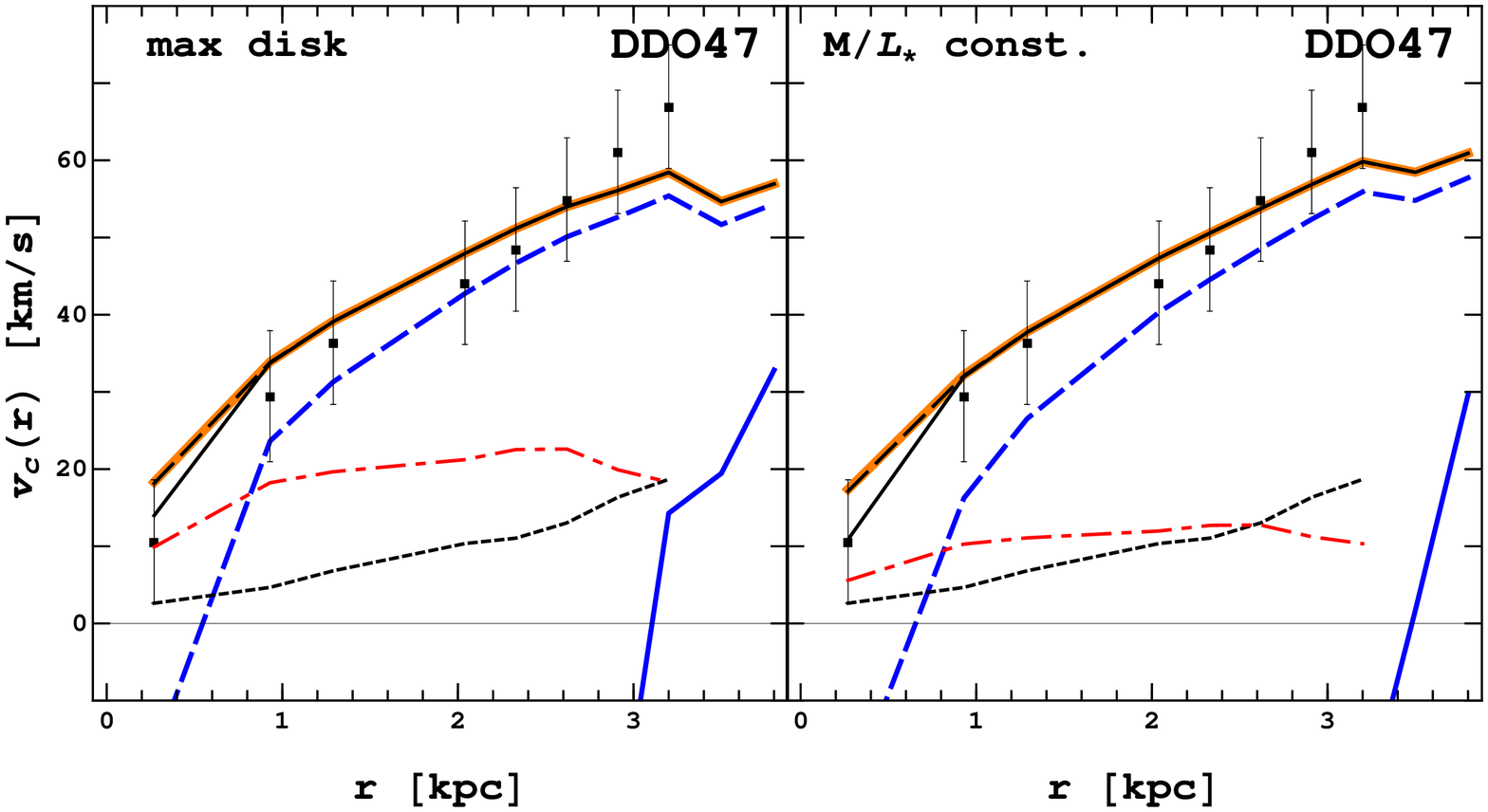} \hfill
\includegraphics[width=8.8cm]{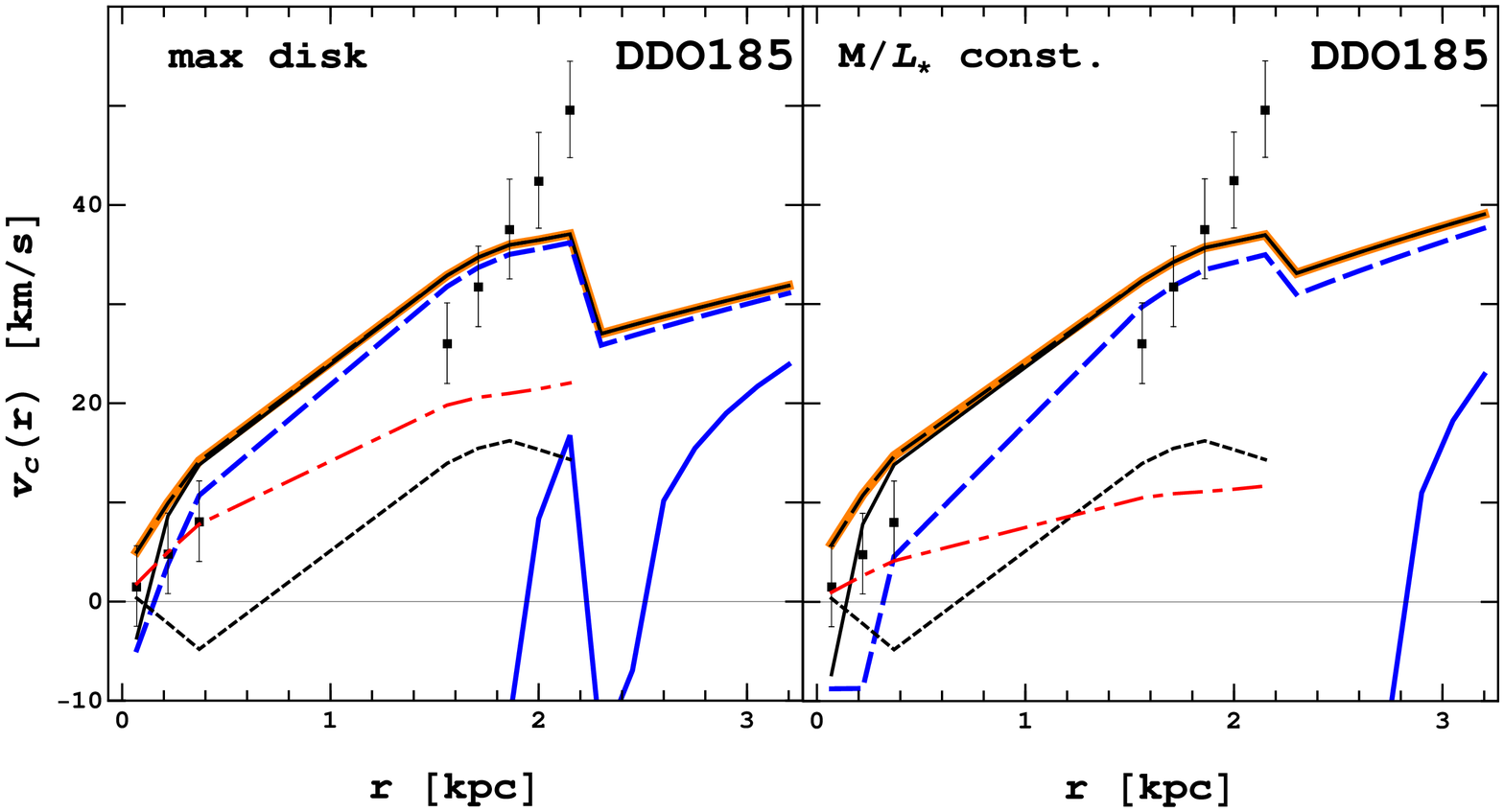}
\caption{Continued.} \label{figGalaxies2}
\end{figure*}
%__________________________________

Depending on the model parameters, the acceleration caused by the
electron-positron gas may be comparable to (or even higher than) the
gravitational one in the central parts of the halo. For our canonical model
(left panel of Figure~\ref{figAcceleration}), the pressure gradient
is strong enough to overcome gravity for $E_0 < 1$~GeV, and the radius at which
both forces balance each other is of the order of 100~pc. The
effect of dark matter annihilation is weaker, but perhaps still measurable, for
$E_0 \sim 1$~GeV. It would be extremely difficult to detect at
10~GeV, and completely negligible for larger particle masses. These conclusions
are very robust with respect to variations in the ionization
fraction of the gas or the intensity of the magnetic field. The exact density of
the interstellar medium has a somewhat larger influence on the
results, partly because of its effect on the dark matter pressure (see
Figure~\ref{figAstro}) and partly through the presence of the gas density
in equation~(\ref{eqAcceleration}). For the extreme case $E_0=100$~MeV and
$\rho_{\rm g}/m_{\rm p}=0.1$~cm$^{-3}$, dark matter pressure is able
to prevent gravitational collapse within the inner 2~kpc, compared to 100~pc for
a density of 10~cm$^{-3}$. However, the qualitative picture is
not changed. For $E_0>10$~GeV, the gravitational acceleration dominates by
several orders of magnitude at all radii, even for the most dilute
gas.

As shown in the previous section, the logarithmic slope of the density profile
plays a critical role on the pressure profile.
The accelerations for $\alpha=1.5$ and 1.9 are shown on the centre and right
panels of Figure~\ref{figAcceleration}, respectively.
When $\alpha=1.9$, the radius at which the relativistic pressure balances
gravity ranges from a few tens of parsec up to several kpc, and a sizable
effect on the rotation curve of the Galaxy is expected for any value of the
injection energy $E_0 < 1$~TeV.

Figure~\ref{figVelocity} shows the modified circular velocity profiles,
according to expression~(\ref{eqVelocity}), for different values of $\alpha$ and
$E_0$.
In our canonical model, the rotation curve of the Galaxy changes significantly
for $E_0 = 100$~MeV, it is slightly modified for $E_0 = 10$~MeV and almost
imperceptibly for $E_0=1$~GeV.
For higher values of the inner logarithmic slope, as those predicted, for
instance, in the adiabatic contraction scenario \citep{Blumenthal+86}, it is
more likely that the annihilation of dark matter particles leaves a clearly
detectable imprint on the observed rotation curve.
The scales on which such a signal would be visible, of the order of kpc in some
cases, subtend several degrees on the sky for the Milky way, and may be
observable as well in other nearby galaxies.

\Referee
{
In other to quantify the effect on the rotation curves of low surface
brightness galaxies, we compute the model predictions for the objects compiled
by \citet{deBlokBosma02}.
Since it is not our aim to fit the data (which would require more careful
modelling, beyond the scope of the present work), we simply take the observed
rotation curves, as well as the quoted decomposition into stellar disk, gaseous
disk, and dark matter halo.
We consider their constant mass-to-light ratio and maximum disk models, adopting
the corresponding best-fitting values of $V_{200}$ and $c_{200}$ \citep[Table~4
in][]{deBlokBosma02}.
Values of $\rho_{\rm s}$ and $r_{\rm s}$ for the maximum disk and constant
mass-to-light ratio cases are given in Tables~\ref{tabGalaxies_maxDisk}
and~\ref{tabGalaxies_ML}, respectively.
}

\Referee
{
Figures~\ref{figGalaxies1} and~\ref{figGalaxies2} show the predicted rotation
curves for $\gamma_0=200$ ($m_{\rm dm} c^2 \sim 100$~MeV) and $\gamma_0=2000$
($m_{\rm dm} c^2 \sim 1$~GeV) with $\langle\sigma v \rangle_{e^\pm} = 2.6 \times
10^{-26}\ {\rm cm^3\ s^{-1}}$ and all other parameters set to our canonical
values.
}
\Referee
{
Since these rotation curves have been computed (at least, in the innermost
regions) from the H$\alpha$ line, we also plot the results obtained for a gas
density $n = \rho_{\rm g}/m_{\rm p} = 0.01$~cm$^{-3}$ and $x_{\rm ion} = 1$,
appropriate for the hot, diffuse component responsible for the emission line.
The original model of \citet{deBlokBosma02} without dark matter annihilation is
shown for the sake of comparison, and the reduced $\chi^2$ values associated to
each model are listed in Tables~\ref{tabGalaxies_maxDisk}
and~\ref{tabGalaxies_ML}.
}

\Referee
{
In general, the effect is not very significant for $n=1$~cm$^{-3}$ (not even for
$m_{\rm dm} c^2 \sim 100$~MeV, except for a few exceptional cases, such as
NGC\,3274).
For the adopted value of the logarithmic slope of the dark matter density
profile near the centre, $\alpha = 1$, and the extremely low values of the
characteristic density $\rho_{\rm s}$ reported by \citet{deBlokBosma02}, the
circular velocity at the innermost point becomes reduced by an amount that is
typically much smaller than the observational error bars.
A more noticeable effect would be obtained for steeper profiles (see
Figure~\ref{figVelocity}), but also if one considers the typical density of the
hot, ionized medium where the H$\alpha$ line originates.
Using $n= 0.01$~cm$^{-3}$, the rotation curves of \emph{all} galaxies would be
dramatically affected on $\sim$~kpc scales for $\alpha = 1$ and $E_0\le$~GeV,
both for the constant $M/L$ and maximum disk models.

These results represent a double-edged sword for dark matter annihilation
models.
On the one hand, it might be possible to find a particular dark matter candidate
that is able to explain the rotation curve data with a cuspy density profile.
On the other hand, we also predict that, in that case, one should observe
prominent differences in the kinematics of the stellar, neutral, and ionized
components due to their different densities.
The observed rotation curves provide thus an additional tool (complementary to
radio and gamma-ray constraints) to rule out a broad class of models and
hopefully help to identify the physical properties of dark matter particles.
}

%--------------------------------------------------------------------------
 \section{Conclusions}
 \label{Conclusions}
%--------------------------------------------------------------------------

In this paper, we have investigated the contribution to the total gas pressure
arising from relativistic electrons and positrons produced in
dark matter annihilations. The propagation of these particles through the ISM is
determined by the diffusion-loss equation. We assume a uniform
diffusion coefficient and consider inverse Compton scattering, synchrotron
radiation, Coulomb collisions, bremsstrahlung and ionization of
neutral hydrogen atoms as the main energy loss mechanisms. All the electrons and
positrons are injected with an initial energy $E_0$ between
1~MeV and 1~TeV, and the injection rate is constrained by different Galactic
observations.

We have evaluated the effect of this ``dark matter pressure'' for astrophysical
conditions representative of the Milky Way and varied the
adopted values of each parameter (intensity of the magnetic field, density and
ionization fraction of the ISM gas, inner logarithmic slope of
the dark matter density profile and the virial mass of the galaxy) to verify
that our results hold in the general case. Our main conclusions can
be summarized as follows:

\begin{enumerate}

\item For the canonical Milky Way model, the dark matter pressure gradient is
able to offset the gravitational acceleration within the central
$\sim10-400$~pc as long as the injection energy is lower than 1~GeV. There would
be an extremely weak signature if $E_0 \sim$~GeV, and the
effect would be completely negligible for larger values of $E_0$.

\item The ionization fraction of the ISM and the intensity of the magnetic field
determine the energy losses and the shape of the electron-positron spectrum at
low and high values of the Lorentz factor, respectively.
Although these details may have a strong impact on other observables, such as
the emission at different wavelengths, they do not affect the rotation curve
significantly.
The precise value of the gas density plays a more important role,
\Referee{and it changes the results at the quantitative level}.

\item Steep logarithmic slopes of the dark matter density profile yield much
higher pressures in the central regions.
For $\alpha \geq 1.9$, a clear signature of dark matter annihilation on the
observed rotation curve is expected even for $E_0 \sim 1$~TeV.

\Referee
{
\item
Comparison with publicly-available observational data shows that, while dark
matter pressure may bring the predicted rotation curves into better agreement
with observations, it is arguably more likely that this effect is more useful as
a constraint on the annihilation cross-section as a function of dark matter
particle mass.
% A more detailed analysis is work in progress.
}

\end{enumerate}

%--------------------------------------------------------------------------
 \section*{Acknowledgments}

%--------------------------------------------------------------------------
M.~Wechakama would like to thank V.~M\"{u}ller and the members of the cosmology
group at the AIP for useful discussions, as well as F.~Breitling
for his help with programming questions and S. White for English corrections.
This work has been funded by DFG Research Grant AS\,312/1-1
(Germany). Y.~Ascasibar also acknowledges support from project AYA
2007-67965-C03-03 (MEC, Spain). We would also like to thank the referee for a very helpful report.

%%%%%%%%%%%%%%%%%%%%%%%%%%%%%%%%%%%%%%%%%%%%%%%%%%%%%%%%%%%%%%%%%%%%%%%%%%%%%%%
 \bibliographystyle{mn2e}
 \bibliography{references}
%%%%%%%%%%%%%%%%%%%%%%%%%%%%%%%%%%%%%%%%%%%%%%%%%%%%%%%%%%%%%%%%%%%%%%%%%%%%%%%

\end{document}